\documentclass[twocolumn]{IEEEtran}
\usepackage[utf8]{inputenc}
\usepackage{textcomp}
\usepackage{lineno}
\usepackage{graphicx,color}
\usepackage{amsmath,amsthm,amssymb}
\usepackage{longtable,tabularx}
\usepackage{subcaption}
\usepackage{cite}

\newtheorem{theorem}{Theorem}
\newtheorem{assumption}{Assumption}
\newtheorem{lemma}{Lemma}

\title{Nonlinear Model Predictive Control Framework For Cooperative Three-Agent Target Defense Game}

\author{Amith Manoharan and P.B. Sujit %
\thanks {Graduate student, Department of Electronics and Communications Engineering, IIIT Delhi, Okhla Phase 3, New Delhi -- 110020.}
\thanks{Associate Professor, Department of Electrical Engineering and Computer Science, Indian Institute of Science Education and Research Bhopal, Bhopal -- 462038.}}

\begin{document}
\maketitle
\IEEEpeerreviewmaketitle
 \begin{abstract}
 This paper presents cooperative target defense guidance strategies using nonlinear model predictive control (NMPC) framework for a target-attacker-defender (TAD) game. 
		The TAD game consists of an attacker and a cooperative target-defender pair. The attacker's objective is to capture the target, whereas the target-defender team acts together such that the defender can intercept the attacker and ensure target survival.  We assume that the cooperative target-defender pair  do not have perfect knowledge of the attacker states, and hence the states  are estimated using an Extended Kalman Filter (EKF). The capture analysis based on the Apollonius circles is performed  to identify the target survival regions. The efficacy of the NMPC-based solution is evaluated through extensive numerical simulations. The results show that the NMPC-based solution offers robustness to the different unknown attacker models and has better performance than CLOS and A-CLOS based strategies.
\end{abstract}

\section{Introduction}\label{sec:intro}
Pursuit-evasion games are useful in several important security applications \cite{bhattacharya2014surveillance}. These games consist of two agents, (i) pursuer agent (P) and (ii) evader agent (E). The pursuer objective is to capture the evader while the evader wants to avoid capture. The most common application of such games is where an aircraft tries to evade an incoming missile. A broad survey on pursuit-evasion differential games is given in \cite{weintraub2020introduction}, and several other studies can be found which analyze games with a single pursuer-evader pair \cite{c6,sunkara2018pursuit,jagat2017nonlinear}, and multiple pursuers and a single evader \cite{c2,sun2017multiple,ramana2017pursuit,pachter2020cooperative}.   

In the presence of multiple agents, cooperation among the agents become important. One such game where there are multiple agents is the three-agent target-attacker-defender (TAD) game. The TAD game consists of an agent named the attacker $(A)$, who pursues to capture the second agent known as the target $(T)$, and the third agent, called the defender $(D)$, tries to help the target by intercepting the attacker before it reaches the target. The target and the defender act as a team, and this cooperation enables the target to maneuver in such a way that the defender would be able to intercept the attacker promptly.    

The TAD problem was introduced by Isaacs in \cite{c19} and was discussed in detail by various researchers \cite{ratnoo2011line,yamasaki2013modified,c8,c11,c12,garcia2018design,garcia2021complete}. Diverse approaches were taken to tackle this problem, including line-of-sight (LOS) guidance \cite{ratnoo2011line,yamasaki2013modified}, linear quadratic regulator \cite{c8,perelman2011cooperative}, and optimal game theoretical solutions \cite{ratnoo2012guidance,garcia2018design,garcia2021complete}. In \cite{c1}, the TAD problem is solved using optimal control theory wherein the optimal heading angles for the target-defender team are determined, assuming that the attacker implements a conventional missile guidance law such as proportional navigation (PN) or pure pursuit guidance (PP). Another optimal control formulation with bounded controls is given in \cite{rubinsky2014three}. In \cite{kumar2017cooperative}, a nonlinear guidance strategy using sliding mode control technique is proposed for various scenarios involving an attacking missile, a target aircraft, and a defender missile. In \cite{yamasaki2010triangle}, the TAD problem is solved using a simple geometrical approach referred to as the triangle guidance, which was also extended to multiple attacker scenarios. The TAD solution using optimal control theory with escape regions for the target is given in \cite{garcia2017cooperative}. Weiss et al. \cite{weiss2017combined} proposed two guidance algorithms for the TAD game. The first one is a combined guidance algorithm for the attacker that simultaneously achieves evasion from the defender and pursuit of the target. The second is a cooperative guidance algorithm for the target-defender pair to enable target escape and the interception of the attacker by the defender. In \cite{harel2020rationalizable}, a three-agent formulation different from the classical TAD game is presented where a navigating aircraft pursues a target and simultaneously tries to avoid an incoming bleeding-energy missile which is assigned to defend the target. In \cite{chipade2019multiplayer}, a multiplayer TAD game is discussed, where Lyapunov-based control strategies are derived for the players using approximations of the minimum and maximum functions. The solutions obtained through analysis of different agent combinations are utilized to pair defenders with attackers to maximize the number of attackers captured. In \cite{manoharan2019nmpc}, the approach does not require perfect knowledge of the states as the agents estimate through the range and heading measurements using an EKF. Manyam et al. \cite{manyam2019coordinating} presented a path planning problem modeled as a TAD game, involving two cooperative agents in an adversarial environment. The second vehicle helps the first vehicle complete a mission by defending against any attackers along the path.  
  
The approaches presented in \cite{ratnoo2011line,yamasaki2013modified,c8,c11,c12,garcia2018design,garcia2021complete} assume that (i) perfect information about attacker states and guidance laws employed by it are known, by which closed-form solutions where developed, and (ii) the target is always at motion, generally with constant speed. In this paper, we relax the above assumptions by using a control scheme that combines the NMPC with attacker state estimation using an EKF. Since the NMPC computes control commands based on the current state estimates, it provides the flexibility to adapt to situations where the attacker can be intelligent. This adaption is not possible in the case of open-loop optimal control formulations where the solution is predetermined \cite{c12,ratnoo2012guidance,rubinsky2014three}. The NMPC also has the advantage of combining optimality with real-time implementability, whereas the conventional missile guidance laws are either sub-optimal or difficult to implement. We also consider the case in which the target has the freedom to move or not to move. This modification is significant for applications like aerial surveillance, communication relaying, etc., where some information needs to be acquired from hostile environments by hovering over the prescribed location. In such situations, the defender can protect the target from any attacking agents while allowing the target to continue its mission. 

The main contributions of this article are (1) NMPC formulation for the TAD game without any assumptions on the attacker states or guidance laws, (2) theoretical analysis of the escape region for the proposed three-agent game for the constant speed target and the variable velocity target, and (3) evaluation of the proposed approach through numerical simulations under different conditions to verify the escape region results, and compare with the CLOS \cite{ratnoo2011line} and A-CLOS \cite{yamasaki2013modified} guidance to show the superiority of the proposed approach. This article significantly extends the article  \cite{manoharan2019nmpc} by providing detailed capturability analysis (contribution (2)), and evaluate the performance satisfying the theoretical properties shown in contribution (2) through simulations under different conditions and show comparison with the \cite{ratnoo2011line} and \cite{yamasaki2013modified}.

The rest of the paper is organized as follows. The problem formulation is given in Section~\ref{sec:probformulation}. The nonlinear model predictive control scheme is explained in Section~\ref{sec:nmpc}. Analysis of the target escape region for the constant speed case and the variable velocity case is given in Section~\ref{sec:escape_region}. Simulation results and the comparison with the CLOS and A-CLOS guidance laws are presented in Section~\ref{sec:results} and conclusions in Section~\ref{sec:conclusions}.   
\section{Problem Formulation}\label{sec:probformulation}
Consider an attacker pursuing a target, while the defender aims to intercept the attacker before $A$ intercepts $T$ as shown in  Fig.~\ref{fig:engagement_trajectory1}. The game is formulated in a 2D Cartesian space, assuming that the altitude of the agents remain constant throughout the engagement. Several assumptions are considered in this article that are listed below.
\subsection{Assumptions}
\begin{assumption}
The agents have point mass, and hence we consider the kinematic equations only.
\end{assumption}
\begin{assumption}
The target-defender team does not have information about the attacker states. They are estimated.
\end{assumption}
\begin{assumption}
The attacker has the capability to acquire the state information of $T$ and $D$ agents accurately.
\end{assumption}
\begin{assumption}
The process and measurement noises $q(k) $ and $ \mu(k) $ are additive and zero mean white Gaussian.
\end{assumption}
\begin{assumption}
The target-defender team has the capability to measure the range and the LOS angle using on-board sensors. \label{assump:sensor}
\end{assumption}

\subsection{System model}
The equations of motion for the agents are given as 
\begin{eqnarray}
\dot{x}_A\left( t\right) &=&v_A\cos{\alpha_A\left( t\right)}, \\
\dot{y}_A\left( t\right)&=&v_A\sin\alpha_A\left( t\right), \\
\dot{x}_T\left( t\right)&=&v_T\left( t\right)\cos\alpha_T\left( t\right),\label{targetstatex}\\
\dot{y}_T\left( t\right)&=&v_T\left( t\right)\sin\alpha_T\left( t\right),\label{targetstatey} \\
\dot{x}_D\left( t\right)&=&v_D\cos\alpha_D\left( t\right), \\
\dot{y}_D\left( t\right)&=&v_D\sin\alpha_D\left( t\right), 
\end{eqnarray}
where $ v_A,~v_T\left( t\right)$ and $v_D $ are the velocities of the attacker, the target, and the defender, respectively. $ v_A,~v_D $ are assumed to be constant, and $ v_T\left( t\right)  $ is bounded by $ v_T\left( t\right) \in [0,\bar{v}_T] $. Similarly $\alpha_A\left( t\right),~\alpha_T\left( t\right)$ and $\alpha_D\left( t\right)  $ are the heading angles of the attacker, the target and the defender respectively. All the states, heading angles of the agents, and the velocity of the target are changing with respect to time $t$, and the notation $ \left( t\right)  $ is omitted in the rest of the paper for simplicity. Previous studies~\cite{c1,c12} assumes that the attacker uses a predefined guidance law. However, an intelligent attacker can use guidance laws which are unknown and hence, we assume that the exact attacker guidance law is not known to the target-defender team. 
\subsection{Engagement Geometry} 
\begin{figure}
	\centering
	\includegraphics[width=7cm]{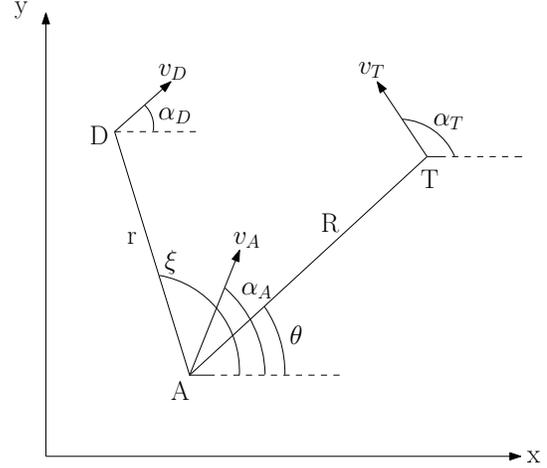}
	\caption{Attacker-Target-Defender engagement geometry.}
	\label{fig:engagement_trajectory1}
\end{figure}
\begin{figure}
	\centering
	\includegraphics[width=7cm]{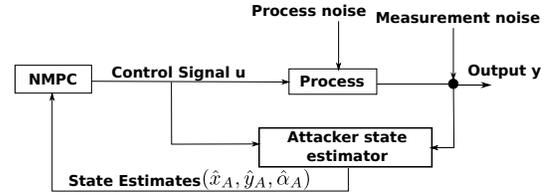}
	\caption{NMPC scheme for determining the control commands for the agents. }
	\label{fig:mpc_block_diagram}
\end{figure}
Consider the three-agent engagement geometry as shown in Fig.~\ref{fig:engagement_trajectory1}. The relative motion of the three agents can be modeled as
\begin{eqnarray}
v_{R}&=&\dot{R}=v_{T}\cos\left(\alpha_{T}-\theta\right) - v_{A}\cos\left(\alpha_{A}-\theta\right),\\
v_{\theta}&=&R\dot{\theta}=v_{T}\sin\left(\alpha_{T}-\theta\right) - v_{A}\sin\left(\alpha_{A}-\theta\right),\\
v_{r}&=&\dot{r}=v_{D}\cos\left(\alpha_{D}-\xi\right) - v_{A}\cos\left(\alpha_{A}-\xi\right),\\
v_{\xi}&=&r\dot{\xi}=v_{D}\sin\left(\alpha_{D}-\xi\right) - v_{A}\sin\left(\alpha_{A}-\xi\right),\\
R&=&\sqrt{\left( x_{T}-x_{A}\right) ^{2}+\left( y_{T}-y_{A}\right) ^{2}},\\
\theta&=&\arctan \left( \frac{y_{T}-y_{A}}{x_{T}-x_{A}}\right),\\ 
r&=&\sqrt{\left( x_{D}-x_{A}\right) ^{2}+\left( y_{D}-y_{A}\right) ^{2}},\\
\xi&=&\arctan \left( \frac{y_{D}-y_{A}}{x_{D}-x_{A}}\right),
\end{eqnarray}
where $\theta$ is the line-of-sight (LOS) angle between the target and the attacker, $ \xi $ is the LOS angle between the defender and the attacker, $R$ is the distance between the attacker and the target, and $r$ is the distance between the defender and the attacker. $v_R$ and $v_{\theta}$ are the components of the relative velocity along the $ A-T $ LOS and perpendicular to the $ A-T $ LOS. $v_r$ and $v_{\xi}$ are the components of the relative velocity along the $ A-D $ LOS and perpendicular to the $ A-D $ LOS.

\section{Nonlinear model predictive control design} \label{sec:nmpc}
Nonlinear model predictive control (NMPC) is based on determining optimal future actions of an agent or a system for a finite time horizon. The main component of the NMPC framework is the mathematical model of the three-agent system~\eqref{eqn:nmpc_model}. For each time step $ t $, the optimal control inputs are computed using this model for the horizon $ [t,t+\tau_{h}]$, where $ \tau_{h} $ is the specified look-ahead window. After applying the first value from the computed control sequence to the real system, the resulting output from the system is feed-backed to the controller, and the process is continued \cite{c15}. We propose to use NMPC to compute the cooperative control commands for the target-defender team. The attacker states are estimated with the help of measurements available to the $T-D$ team. Fig.~\ref{fig:mpc_block_diagram} shows the proposed NMPC strategy, which also includes the state estimator for the attacker in the feedback loop in addition to the general NMPC structure. The NMPC block is the core component of the scheme, which contains the mathematical model of the system and a numerical optimizer. Process block represents the system involving the agents, and the estimator block contains the EKF.

\subsection{Objective function}
We propose a strategy wherein the target-defender team uses NMPC to compute the control commands so that the objective of target evasion from the attacker while the defender intercepts the attacker is achieved. We consider two types of target maneuvers: 1) constant speed target with only heading change, and 2) variable velocity target.
The variable velocity target implies that the target can move or stop based on predefined conditions.   

\begin{figure}
	\centering
	\includegraphics[width=6cm]{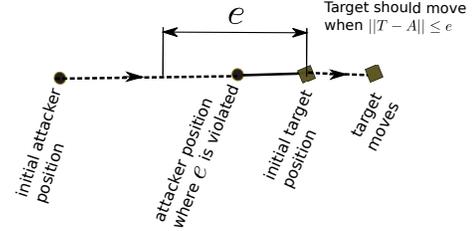}
	\caption{Definition of safe distance of the target from the attacker. }
	\label{fig:e}
\end{figure}

The NMPC problem can be stated as:
\begin{align}
&\min_{\boldsymbol{u_{x},u_{y},\dot{\alpha}_{D}} \in \mathcal{PC}(t,t+\tau_h)}J(X,U,t)\nonumber\\
&=\int_{t}^{t+\tau_{h}}\left[  \left( u_{x}^{2}+u_{y}^{2}\right)+r+\max(0,e-R)\right]  \mathop{}\!\mathrm{d}t,
\end{align}
subject to:
\begin{eqnarray}\label{eqn:nmpc_model}
\dot{X}&=&f\left( X,U,t\right),\\
U&\in&\left[ U^{-},U^{+}\right], \\
\sqrt{u_x^2+u_y^2} &\leq& \bar{v}_T,
\end{eqnarray}
where $ X=\{ x_{A},y_{A},x_{T},y_{T},x_{D},y_{D},R,\theta\} $ are the states, $ U=\left\lbrace u_{x},u_{y},\dot{\alpha}_{D}\right\rbrace  $ are the control inputs, $ U^{-} $ and $ U^{+} $ are the lower and upper bounds of $ U$, $ \mathcal{PC}(t,t+\tau_h) $ denotes the space of piece-wise continuous function defined over the time interval $ \left[ t,t+\tau_h\right]  $, and the target velocity is bounded by $\bar{v}_T$. We define a threshold, $ {e} $, which is the safe distance of the target from the attacker as shown in Fig.~\ref{fig:e}. The target should move only if this safe distance is violated. For including the variable velocity characteristic of the target, the state equations (\ref{targetstatex}) and (\ref{targetstatey}) are modified as 
\begin{eqnarray}
\dot{x}_{T}&=&u_{x},\\
\dot{y}_{T}&=&u_{y},
\end{eqnarray}  
where $ u_{x},u_{y} $ are the velocity components of the target in $x$ and $y$ directions, respectively, which will also act as the control inputs for the target. The target heading angle $\alpha_{T} $ can be written in terms of the control inputs as
\begin{equation}
\alpha_{T}=\arctan \left( \frac{u_{y}}{u_{x} }\right).
\end{equation} 
The attacker states are unknown and we estimate these states using an extended Kalman filter (EKF).
\subsection{Estimation of the attacker states using EKF}\label{sec:ekf}
An extended Kalman filter (EKF) is used to estimate the attacker's position and course angle. The EKF is formulated as~\cite{c18},

\noindent{Model}
\begin{eqnarray}
X_A(k)&=&f_A\left( X_A(k-1),U(k),k\right) +q(k),\\
z(k)&=&h\left( X_A(k),U(k),k\right) +\mu(k),
\end{eqnarray}  
\noindent {Prediction}
{\small
\begin{align}
X_A(k|k-1)&=f_A( X_A(k-1|k-1),U(k),k)\Delta t  \\ &+X_A(k-1|k-1), \nonumber \\
P(k|k-1)&=\nabla F_{X_A}P(k-1|k-1)\nabla F_{X_A}'+Q,\\
z(k|k-1)&=h( X_A(k|k-1)),
\end{align}
}
\noindent {Update}
{\small
\begin{eqnarray}
X_A(k|k)&=&X_A(k|k-1)+K(k)\nu(k),  \\
P(k|k)&=&P(k|k-1)-K(k)S(k)K'(k),\\
\nu(k)&=&z(k)-z(k|k-1),\\
K(k)&=&P(k|k-1)\nabla H_{X_A}'S^{-1}(k),\\
S(k)&=&\nabla H_{X_A}P(k|k-1)\nabla H_{X_A}'+\Sigma,
\end{eqnarray}}
where, ${X_A}(k)$ and $z(k) $ represent the attacker state model and measurement model respectively, and $P(k)$ is the covariance matrix. $ q(k) $ and $ \mu(k) $ are the process and measurement noises with covariance $ Q $ and $ \Sigma $ respectively. $ \nu(k) $ is called as the innovation parameter, and $ K(k) $ is the Kalman gain. The attacker states $\left\lbrace x_A,y_A,\alpha_A\right\rbrace $ and control $ a_A $ need to be estimated, hence the prediction model is defined as:
\begin{equation}
f_A=\begin{bmatrix}
\dot{x}_{A}\\
\dot{y}_{A}\\
\dot{\alpha}_{A}\\
\dot{a}_A
\end{bmatrix}=
\begin{bmatrix}
v_{A}\cos\alpha_{A}\\
v_{A}\sin\alpha_{A}\\
\frac{a_{A}}{v_A}\\
-a_A
\end{bmatrix},
\end{equation}
and the Jacobian of $ f_A $ is
\begin{equation}
\nabla F_{X_A}=
\begin{bmatrix}
0 & 0 & -v_{A}\sin\alpha_{A} & 0\\
0 & 0 & v_{A}\cos\alpha_{A} & 0\\
0 & 0 & 0 &\frac{1}{v_A}\\
0 & 0 & 0 & -1
\end{bmatrix}.
\end{equation} 
According to Assumption~\ref{assump:sensor}, the measurement model for $ R,r,\theta $ and $ \xi $ is given as 
\begin{equation}
h = \begin{bmatrix}
\sqrt{\left( x_{T}-x_{A}\right) ^{2}+\left( y_{T}-y_{A}\right) ^{2}}\\
\sqrt{\left( x_{D}-x_{A}\right) ^{2}+\left( y_{D}-y_{A}\right) ^{2}}\\
\tan^{-1} \left( \frac{y_{T}-y_{A}}{x_{T}-x_{A}}\right)\\
\tan^{-1} \left( \frac{y_{D}-y_{A}}{x_{D}-x_{A}}\right)
\end{bmatrix}.
\end{equation}
The Jacobian of the measurement model is given by
{\small
\begin{align}
&\nabla H_{X_A}= \nonumber\\ &\begin{bmatrix}
\frac{-(x_{T}-x_{A})}{\sqrt{\left( x_{T}-x_{A}\right) ^{2}+\left( y_{T}-y_{A}\right) ^{2}}}&
\frac{-(y_{T}-y_{A})}{\sqrt{\left( x_{T}-x_{A}\right) ^{2}+\left( y_{T}-y_{A}\right) ^{2}}}& 0 & 0\\
\frac{-(x_{D}-x_{A})}{\sqrt{\left( x_{D}-x_{A}\right) ^{2}+\left( y_{D}-y_{A}\right) ^{2}}}&
\frac{-(y_{D}-y_{A})}{\sqrt{\left( x_{D}-x_{A}\right) ^{2}+\left( y_{D}-y_{A}\right) ^{2}}}& 0 & 0\\
\frac{y_{T}-y_{A}}{\left( x_{T}-x_{A}\right) ^{2}+\left( y_{T}-y_{A}\right) ^{2}}&
\frac{-(x_{T}-x_{A})}{\left( x_{T}-x_{A}\right) ^{2}+\left( y_{T}-y_{A}\right) ^{2}}& 0 & 0\\
\frac{y_{D}-y_{A}}{\left( x_{D}-x_{A}\right) ^{2}+\left( y_{D}-y_{A}\right) ^{2}}&
\frac{-(x_{D}-x_{A})}{\left( x_{D}-x_{A}\right) ^{2}+\left( y_{D}-y_{A}\right) ^{2}}& 0 & 0
\end{bmatrix}.
\end{align}
}

\section{Escape region}\label{sec:escape_region}
The ability to capture or escape depends on the initial conditions of the engagement geometry. We present an analysis based on the Apollonius circle concept to determine the escape and capture zones for the target-defender team. 
 Initially we will assume that the attacker--defender agents have equal speed and then relax with assumption to show how the zones are modified. 
\subsection{Assumptions}
\begin{assumption}
The target is slower than the attacker. Otherwise, the target will always evade the attacker, and the defender's role will become insignificant.
\end{assumption}
\begin{assumption}
	The speed of the defender is greater than or equal to the speed of the target. This is a valid assumption since, in real-world scenarios, the target is usually a slow-moving aircraft, and the defender is a missile.
\end{assumption}

\subsection{Constant speed target} \label{sec:CS_target_escape}
\begin{figure}
	\centering
	\includegraphics[width=6cm]{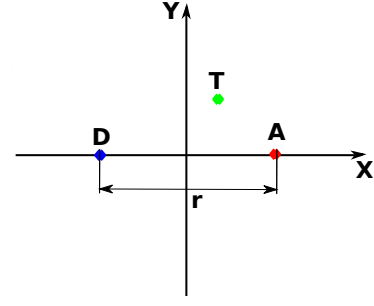}
	\caption{Modified TAD reference frame. }
	\label{fig:tad_frame1}
\end{figure}

We consider a modified reference frame as shown in Fig.~\ref{fig:tad_frame1}, where $ T,A,D $ represents the positions of the target, the attacker, and the defender respectively. The $x-$axis is defined as the line joining the attacker and the defender, $ A $ and $ D $. The $y-$axis is defined as the perpendicular bisector of the line segment $ \overline{AD} $.  

\begin{figure}
	\centering
	\includegraphics[width=6cm]{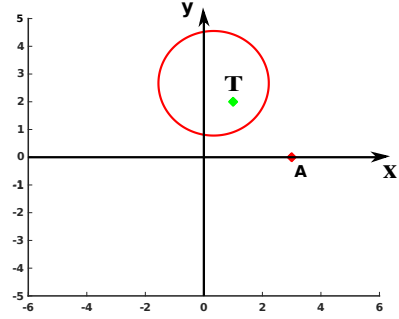}
	\caption{Example Apollonius circle for $ A-T $ engagement. }
	\label{fig:AT}
\end{figure} 

\begin{lemma}\label{lemma:y_axis}
The Apollonius circle for the modified reference frame of the $ A-D $ engagement geometry is the $y-$axis~\cite{c19}.
\end{lemma}
\proof
    The modified reference frame for the $A-D$ engagement is shown in Fig. \ref{fig:tad_frame1}. The center of the Apollonius circle formed by $ A-D $ is 
\begin{equation}
\left( x_c,y_c\right) _{AD} = \left(  \frac{x_{D}-\gamma_{AD}^2x_{A}}{1-\gamma_{AD}^2},\frac{y_{D}-\gamma_{AD}^2y_{A}}{1-\gamma_{AD}^2}\right), 
\label{eq:A_D_circle}
\end{equation}
where $ \gamma_{AD} $ is the speed ratio of the defender to the attacker, $\gamma_{AD} = \frac{v_{D}}{v_{A}}$. 
Since the $x-$axis is defined as the line joining $A$ and $D$, $y_A$ and $y_D$ will always be zero. Therefore, equation~(\ref{eq:A_D_circle}) becomes
\begin{equation}
\left( x_c,y_c\right) _{AD} = \left(  \frac{x_{D}-\gamma_{AD}^2x_{A}}{1-\gamma_{AD}^2},0\right). \label{c_AD}
\end{equation}  
The radius of the Apollonius circle for $A-D$ engagement geometry is given by
\begin{equation}
r_{AD} = \frac{\gamma_{AD} \left( x_{D}-x_{A}\right) }{1-\gamma_{AD}^2}. \label{r_AD}
\end{equation}
Since the attacker and the defender have equal speed, $ \gamma_{AD} = 1 $. According to equations (\ref{c_AD})-(\ref{r_AD}), the Apollonius circle for the $ A-D $ case will become the $y-$axis in Fig.~\ref{fig:tad_frame1}. \hfill $\Box$

\begin{lemma}\label{lemma:a_t_condition}
The defender would be able to intercept the attacker only if the $ A-T $ circle intercepts the $ y- $axis.
\end{lemma}
\proof
Consider the Apollonius circle of $ A-T $ engagement. The center and radius of the Apollonius circle for $A-T$ engagement geometry is given as
\begin{equation}
\left( x_c,y_c\right) _{AT} = \left(  \frac{x_{T}-\gamma_{AT}^2x_{A}}{1-\gamma_{AT}^2},\frac{y_{T}}{1-\gamma_{AT}^2}\right), \label{c_AT}
\end{equation}
and
\begin{equation}
r_{AT} = \frac{\gamma_{AT}\sqrt{ \left( x_{T}-x_{A}\right)^2 +y_{T}^2}}{1-\gamma_{AT}^2}. \label{r_AT}
\end{equation}
Fig.~\ref{fig:AT} represents the Apollonius circle for the $ A-T $ engagement. The circle represents the points at which the attacker and the target can reach simultaneously, leading to the target capture. Since the $ y- $axis represents the $ A-D $ Apollonius circle, the defender would be able to intercept the attacker only if the $ A-T $ circle intercepts the $ y- $axis. \hfill $\Box$

It is possible to map the target escape zone by examining if the condition given in Lemma.~\ref{lemma:a_t_condition} is satisfied or not. In the following theorem, we find the expression for the locus of points that divides the Cartesian plane into the target escape and capture zones.    
\begin{theorem}
	The escape region for the target can be represented by a curve given by
	\begin{equation}
	\frac{x^2}{\gamma_{AT}^2x_{A}^2}-\frac{y^2}{\left( 1-\gamma_{AT}^2\right) x_{A}^2} = 1, \label{eq:equal_hyperbola}
	\end{equation}
	that divides the Cartesian plane into two zones, $ Z_e $ and $ Z_c $, where $ Z_e $ is the escape zone for the target and $ Z_c $ is the capture zone for the target \cite{garcia2018design}.   
\end{theorem}
\proof
	The defender would be able to intercept the attacker before it captures the target only if the $ A-T $ Apollonius circle intercepts the $A-D$ Apollonius circle, which is the $y-$axis, according to Lemma.~\ref{lemma:y_axis} and Lemma.~\ref{lemma:a_t_condition}. The radius of the $A-T$ circle is given by (\ref{r_AT}), and the $x-$coordinate of its center is 
	\begin{equation}
	x_{c_{AT}} = \frac{1}{1-\gamma_{AT}^2}\left( x_T - \gamma_{AT}^2 x_A\right). 
	\end{equation} 
	For the $ A-T $ circle to intercept the $y-$axis, $ r_{AT} $ should be greater than $ x_{c_{AT}} $.
	\begin{align}
	\frac{\gamma_{AT}\sqrt{ \left( x_{T}-x_{A}\right)^2 +y_{T}^2}}{1-\gamma_{AT}^2} &> \frac{1}{1-\gamma_{AT}^2}\left( x_T - \gamma_{AT}^2 x_A\right),\\
	x_T - \gamma_{AT}^2 x_A &< \gamma_{AT}\sqrt{ \left( x_{T}-x_{A}\right)^2 +y_{T}^2},\\
	\frac{x_A^2}{\left(\frac{x_T}{\gamma_{AT}} \right)^2 } + \frac{y_T^2}{\left(\frac{\sqrt{1-\gamma_{AT}^2}}{\gamma_{AT}}x_T \right) ^2} &> 1. \label{eq:inequal_hyperbola}
	\end{align}
	To obtain the curve that defines the escape zone, we equate (\ref{eq:inequal_hyperbola}) to 1. The modified equation is
	\begin{equation}
		\frac{x_A^2}{\left(\frac{x_T}{\gamma_{AT}} \right)^2 } + \frac{y_T^2}{\left(\frac{\sqrt{1-\gamma_{AT}^2}}{\gamma_{AT}}x_T \right) ^2} = 1,
		\end{equation}
	substituting $x_T,y_T$ with $x,y$, we get the equation for the locus of points that divides the plane as
		\begin{equation}
		\frac{x_A^2}{\left(\frac{x}{\gamma_{AT}} \right)^2 } + \frac{y^2}{\left(\frac{\sqrt{1-\gamma_{AT}^2}}{\gamma_{AT}}x \right) ^2} = 1,
		\end{equation}
	which can be rearranged into
	\begin{equation}
	\frac{x^2}{\gamma_{AT}^2x_{A}^2}-\frac{y^2}{\left( 1-\gamma_{AT}^2\right) x_{A}^2} = 1.
	\end{equation} \hfill $\Box$
 
\begin{figure}
	\centering
	\includegraphics[width=6cm]{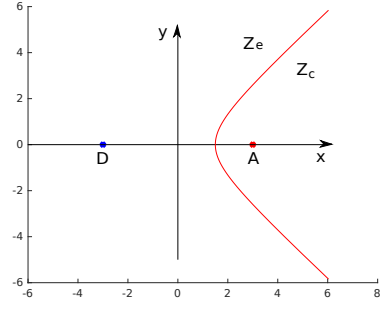}
	\caption{Escape zone for the constant speed target, $T$. $Z_e$ is the escape zone and $Z_c$ is the capture zone. }
	\label{fig:tad_hyper1}
\end{figure}
\begin{figure}
	\centering
	\includegraphics[width=6cm]{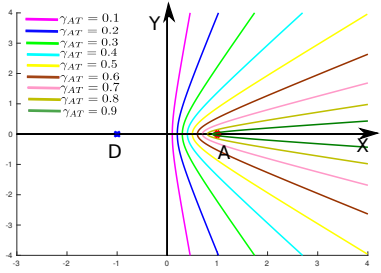}
	\caption{Family of dividing curves for $\gamma_{AT}$ = 0.1,\ldots,0.9. }
	\label{fig:tad_gamma}
\end{figure}

 
Fig.~\ref{fig:tad_hyper1} shows the curve dividing the plane into two zones. If the target's initial position lies in the escape zone $ Z_e $, the $A-T$ Apollonius circle will intercept the $y- $axis and hence, the target can lure the attacker into crossing the $ y- $axis and allow the defender to intercept it. If the target lies in the capture zone $ Z_c $, it would be captured before crossing the $ y- $axis. Fig.~\ref{fig:tad_gamma} shows the family of dividing curves for different values of $\gamma_{AT}$, $0<\gamma_{AT}<1$. It can be seen that as the speed ratio $\gamma_{AT}$ increases, the area of the target escape zone increases since a target with higher velocity has more chance of escaping the attacker.  

\subsection{Stationary target}\label{sec:stationary_target}
\begin{figure}
	\centering
	\includegraphics[width=6cm]{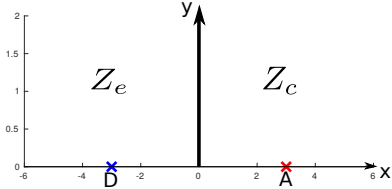}
	\caption{Escape zone for the stationary target, $T$. $Z_e$ is the escape zone and $Z_c$ is the capture zone. }
	\label{fig:stationary_map}
\end{figure}

In the case of a stationary target which does not move throughout the engagement, the $A-T$ speed ratio $\gamma_{AT}=0$. Hence, the center and radius of the $A-T$ Apollonius circle given by equations (\ref{c_AT}) and (\ref{r_AT}) can be modified as
\begin{equation}
\left( x_c,y_c\right) _{AT} = \left(  x_{T},y_{T}\right), \label{eq:stationary_target_center}
\end{equation}
and
\begin{equation}
r_{AT} = 0, \label{eq:stationary_target_radius}
\end{equation}
which means the Apollonius circle will shrink to become the target point itself. According to Lemma.~\ref{lemma:a_t_condition}, the target would be able to escape only if the $A-T$ Apollonius circle intercepts the $y-$axis. Hence, applying the condition for the $A-T$ circle to intercept the $y-$axis, which is $r_{AT}>x_{c_{AT}}$, we get
\begin{equation}
    0>x_{T},
\end{equation}
and after substituting $x$ for $x_T$ and changing the inequality sign to equality for obtaining the dividing curve, we get
\begin{equation}
    x = 0,
\end{equation}
which is the equation for the $y-$axis. Hence, for the stationary target case, the curve that divides the space into the escape and the capture zones is the $y-$axis itself, as shown in Fig.~\ref{fig:stationary_map}.

\subsection{Variable target velocity}\label{sec:var_target_escape}
\begin{figure*}
	\centering 
	\begin{subfigure}{0.5\textwidth}
	\centering
		\includegraphics[width=7cm]{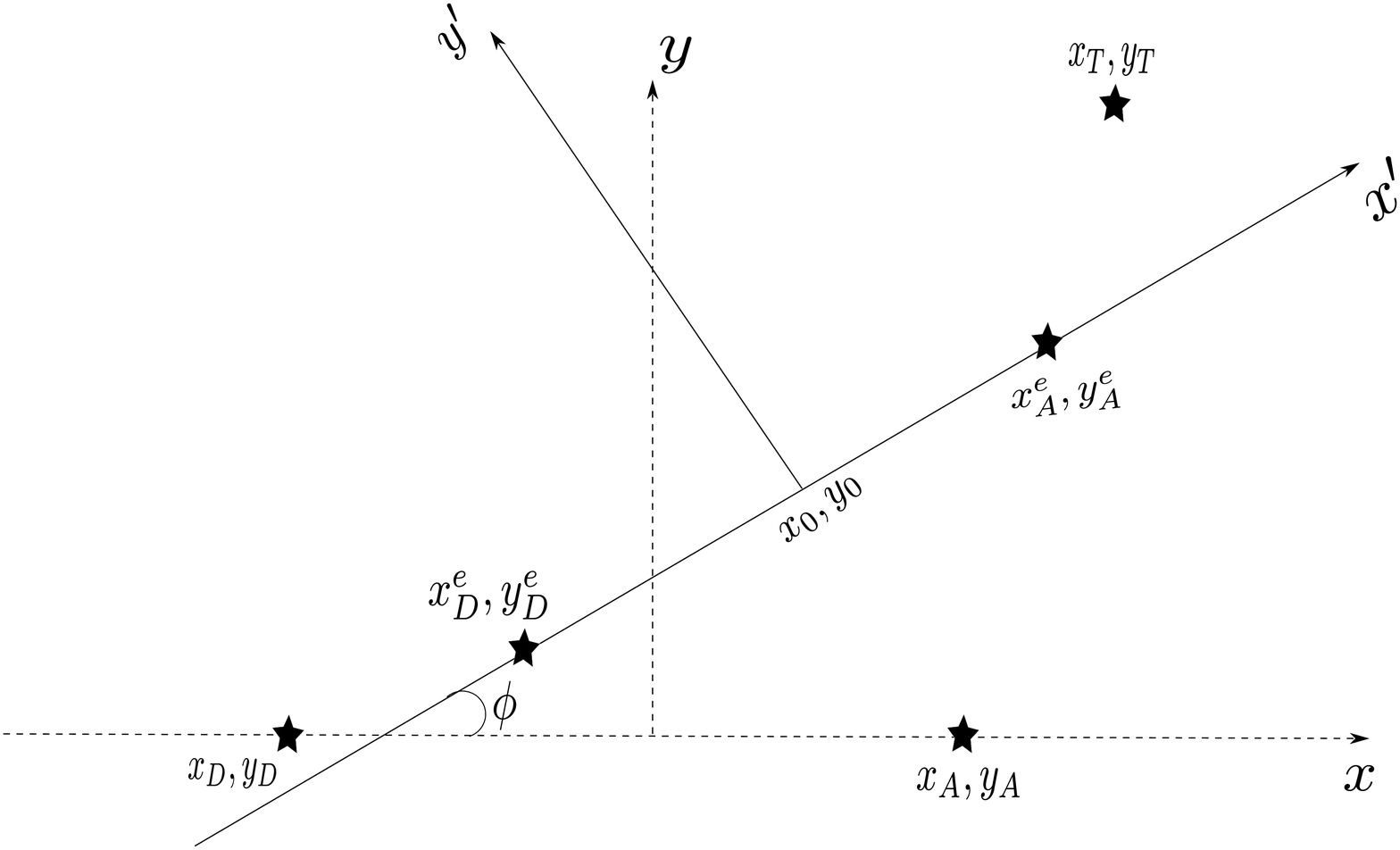}
		\caption{}
		\label{fig:TAD_new_frame}
	\end{subfigure}\hfil 
	\begin{subfigure}{0.5\textwidth}
	\centering
		\includegraphics[width=7cm]{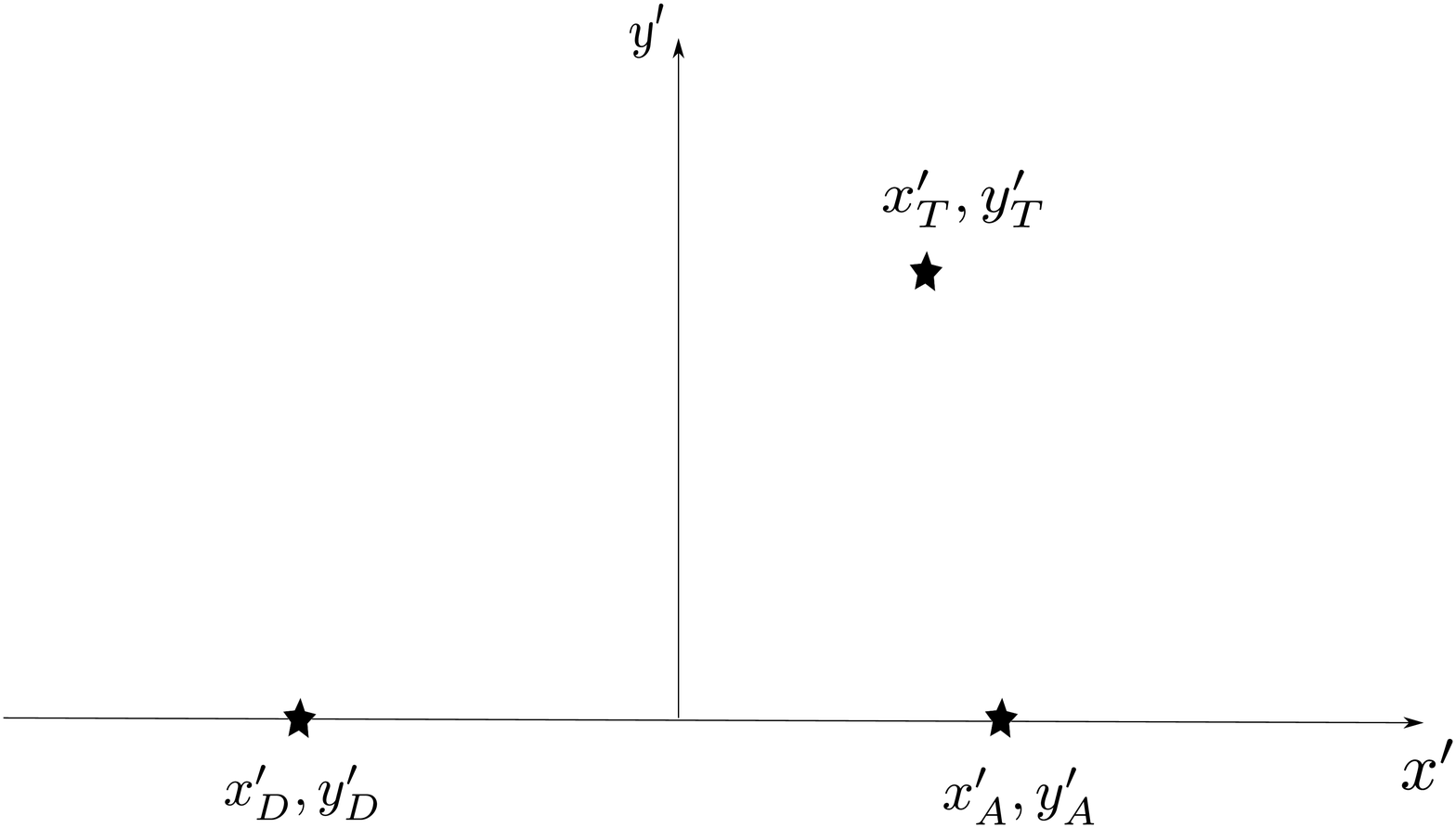}
		\caption{}
		\label{fig:TAD_new_frame_2}
	\end{subfigure}\hfil 
		\caption{Configurations in a new reference frame for the variable velocity target. (a) Modified reference frame (figure not to scale). (b) Agent representations in the new $x'-y'$ frame. }
		\label{fig:TAD_variable_new_frame}
\end{figure*}

For analyzing this case, consider Fig.~\ref{fig:TAD_new_frame}. The target, attacker, and defender initial positions is given as $(x_A,y_A)$, $(x_D,y_D)$, and $(x_T,y_T) $. The target is in motion only when the attacker violates the safe distance $ e $. Let $ (x_A^e,y_A^e) $ and $ (x_D^e,y_D^e) $ be the coordinates of the attacker and the defender when $||T-A|| = e$. We now create a new reference frame $x'-y'$ as shown in the figure, where
\begin{align}
\phi &= \arctan\left(\frac{y_A^e - y_D^e}{x_A^e - x_D^e} \right), \text{and} \\ 
x_0,y_0 &= \left(\frac{x_A^e + x_D^e}{2},\frac{y_A^e + y_D^e}{2} \right). 
\end{align}    
The coordinates from $x-y$ frame can be transformed into $x'-y'$ frame as follows
\begin{align}
x_A' &= \left( x_A^e-x_0\right) \cos\phi + \left( y_A^e-y_0\right) \sin\phi, \\
y_A' &= \left( y_A^e-y_0\right) \cos\phi - \left( x_A^e-x_0\right) \sin\phi, \\
x_D' &= \left( x_D^e-x_0\right) \cos\phi + \left( y_D^e-y_0\right) \sin\phi, \\
y_D' &= \left( y_D^e-y_0\right) \cos\phi - \left( x_D^e-x_0\right) \sin\phi, \\
x_T' &= \left( x_T-x_0\right) \cos\phi + \left( y_T-y_0\right) \sin\phi, \\
y_T' &= \left( y_T-y_0\right) \cos\phi - \left( x_T-x_0\right) \sin\phi.
\end{align}
Fig.~\ref{fig:TAD_new_frame_2} shows the agent representations in the re-defined $ x'-y' $ frame. 

\begin{figure}
	\centering
	\includegraphics[width=8cm]{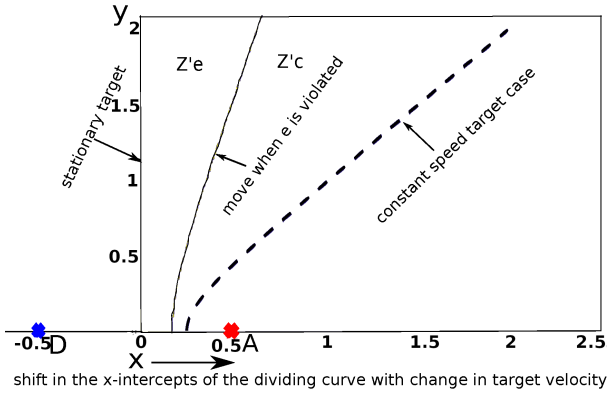}
	\caption{Escape zone for the variable velocity target, $T$. The dashed curve represents the constant speed case, the solid curve the variable velocity case, and the $y-$axis represents the boundary for the stationary target. }
	\label{fig:tad1}
\end{figure}

Consider an example case where the initial conditions are selected as $ (x_A,y_A)=(0.5,0) $, $ (x_D,y_D)=(-0.5,0) $, $ \gamma_{AT}=0.5 $, and $e=\frac{1}{2}R$. From equation (\ref{eq:inequal_hyperbola}), the escape zone for the target is given in Fig.~\ref{fig:tad1}. The variable target velocity case can be subdivided into three sub-cases, (i) if the target is always moving with maximum allowed speed as given in Sec.~\ref{sec:CS_target_escape}. In this case, the dashed curve represents the division of the plane into the escape and the capture zones. (ii) If the target is always stationary as given in Sec.~\ref{sec:stationary_target}, the division will be represented by the $y-$axis. The target escape zone will be left of the $y-$axis and the capture zone will be the right side of $y-$axis. This is due to the fact that the interception of the attacker by the defender can only occur at the $y-$axis and if the target does not move, it has to be in the left side of the $y-$axis in order to lure the attacker into crossing the $y-$axis so that the defender can intercept it. (iii) If the target is initially at rest and starts moving with maximum velocity after the safe distance $e$ is violated. For this case, the solid curve represents the division of the space into the two zones, the target escape zone, $ Z_e' $ and the target capture zone $ Z_c' $. After the target starts moving, the engagement is similar to the constant speed target case given in Sec.~\ref{sec:CS_target_escape}, and the conditions given in Lemma.~\ref{lemma:a_t_condition} can be used to map the escape region for the target. 

Deriving an analytical expression for this solid curve is very difficult, unlike the constant speed case since at each initial target position, a new reference frame needs to be formed, which depends on the instantaneous position of the defender $(x_D^e,y_D^e)$, which is the solution of a differential equation. Hence for the variable target velocity case, the escape and the capture zone are obtained through numerical analysis.  

\subsection{Different velocities for attacker and defender}
Now, we relax the assumption that the attacker and the defender have equal speed and perform the analysis for cases where $\gamma_{AD}>0,\gamma_{AD} \neq 1$.
\subsubsection{Constant speed target case}
Extending Lemma~\ref{lemma:a_t_condition} for general cases, we can see that the curve that separates the escape region from the capture region is the locus of points where the $A-T$ Apollonius circle is tangential to the $A-D$ Apollonius circle. This condition can be written as
\begin{equation}
	d_c = r_{AD}+r_{AT}, \label{eq:condition_gamma_1}
\end{equation}
where $d_c$ is the distance between the centers of the $A-D$ and $A-T$ Apollonius circles, $r_{AD}$ is the radius of the $A-D$ circle, and $r_{AT}$ is the radius of the $A-T$ circle respectively. Now, we substitute and expand the terms in equation~\eqref{eq:condition_gamma_1} as follows
\begin{align}
	&\sqrt{(x_{c_{AD}}-x_{c_{AT}})^2+(y_{c_{AD}}-y_{c_{AT}})^2} = \frac{\gamma_{AD}(x_D-x_A)}{1-\gamma_{AD}^2} \nonumber \\&\qquad+\frac{\gamma_{AT}\sqrt{(x_T-x_A)^2+y_T^2}}{1-\gamma_{AT}^2}, \\
	&\left( \frac{x_D-\gamma_{AD}^2x_A}{1-\gamma_{AD}^2} - \frac{x_T-\gamma_{AT}^2x_A}{1-\gamma_{AT}^2} \right)^2 + \left( \frac{y_T}{1-\gamma_{AT}^2} \right)^2 =\nonumber \\&\qquad \left( \frac{\gamma_{AD}(x_D-x_A)}{1-\gamma_{AD}^2} + \frac{\gamma_{AT}\sqrt{(x_T-x_A)^2+y_T^2}}{1-\gamma_{AT}^2} \right) ^2.
\end{align}  
After some algebraic manipulations and substituting $x,y$ for $x_T,y_T$, and $x_D = -x_A$, we get the following quartic equation
\begin{multline}
	c_3^2y^4+(2c_2c_3x^2+2c_3c_4x+2c_3c_0-c_1^2)y^2 + c_2^2x^4 + c_4^2x^2 + c_0^2 \\+ 2c_4c_0x + 2c_2c_4x^3 + 2c_2c_0x^2 - c_1^2x^2 + 2c_1^2x_Ax-c_1^2x_A^2 = 0, \label{eq:quartic_gamma}
\end{multline}
where
\begin{equation*}
\begin{split}
	c_0 &= x_A^2 - 2\gamma_{AD}^2x_A^2 + \gamma_{AD}^4x_A^2 - 5\gamma_{AT}^2x_A^2 + 8\gamma_{AT}^2\gamma_{AD}^2x_A^2 -
	\\  & \quad \gamma_{AD}^4\gamma_{AT}^2x_A^2 + 4\gamma_{AT}^4x_A^2  -4\gamma_{AT}^4\gamma_{AD}^2x_A^2 - 2\gamma_{AD}^2\gamma_{AT}^2x_A,\\
	c_1 &= -4\gamma_{AT}\gamma_{AD}x_A + 4\gamma_{AT}^3\gamma_{AD}x_A + 4\gamma_{AT}\gamma_{AD}^3x_A -
	\\  & \quad 4\gamma_{AD}^3\gamma_{AT}^3x_A,\\
	c_2 &= 1-\gamma_{AT}^2-2\gamma_{AD}^2+2\gamma_{AT}^2\gamma_{AT}^2+\gamma_{AD}^4-\gamma_{AD}^4\gamma_{AT}^2,\\
	c_3 &= 1-\gamma_{AT}^2-2\gamma_{AD}^2+2\gamma_{AD}^2\gamma_{AT}^2+\gamma_{AD}^4-\gamma_{AD}^4\gamma_{AT}^2,\\
	c_4 &= 2x_A-2\gamma_{AT}^2x_A-2\gamma_{AD}^4x_A+2\gamma_{AT}^2\gamma_{AD}^4x_A.
\end{split}
\end{equation*}
The solutions for the equation~\eqref{eq:quartic_gamma} is given by
\begin{eqnarray}
	y_1^2 &=& \frac{-b+\sqrt{b^2-4ac}}{2a}, \label{eq:gamma_solution1}\\
	y_2^2 &=& \frac{-b-\sqrt{b^2-4ac}}{2a}, \label{eq:gamma_solution2}
\end{eqnarray}
where 
\begin{equation*}
	\begin{split}
		a &= c_3^2,\\
		b &= 2c_2c_3x^2 + 2c_3c_4x + 2c_3c_0 - c_1^2,\\
		c &= c_2^2x^4 + c_4^2x^2 + c_0^2 + 2c_4c_0x + 2c_2c_4x^3 + 2c_2c_0x^2 
		\\  & \quad - c_1^2x^2 + 2c_1^2x_Ax-c_1^2x_A^2.
	\end{split}
\end{equation*}

\begin{figure*}
	\centering 
	\begin{subfigure}{0.5\textwidth}
	\centering
		\includegraphics[width=6cm]{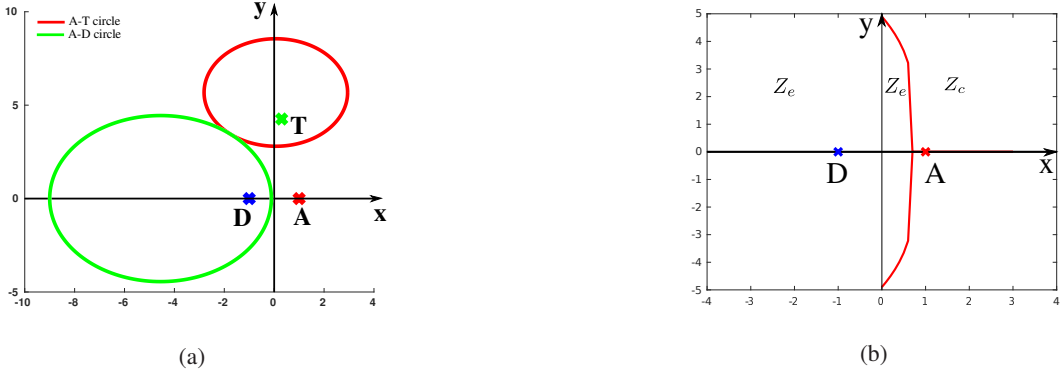}
		\caption{}
		\label{fig:gamma1_circle}
	\end{subfigure}\hfil 
	\begin{subfigure}{0.5\textwidth}
	\centering
		\includegraphics[width=6cm]{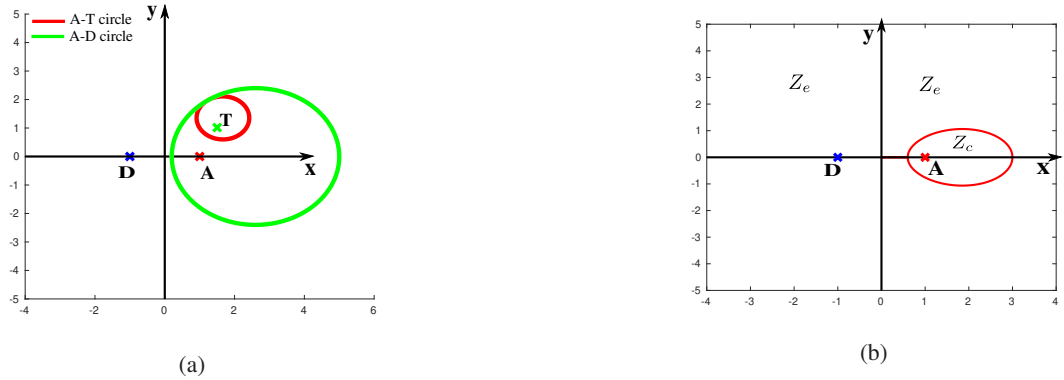}
		\caption{}
		\label{fig:gamma1_escape}
	\end{subfigure}\hfil 
	\caption{Example plots of the escape region for a constant speed target when $\gamma_{AD}<1$. (a) Apollonius circles of $A-D$ and $A-T$ engagements at the target escape boundary. (b) The curve that divides the plane into escape and capture regions for the target. }
	\label{fig:gamma1}
\end{figure*}

\begin{figure*}
	\centering 
	\begin{subfigure}{0.5\textwidth}
	\centering
		\includegraphics[width=6cm]{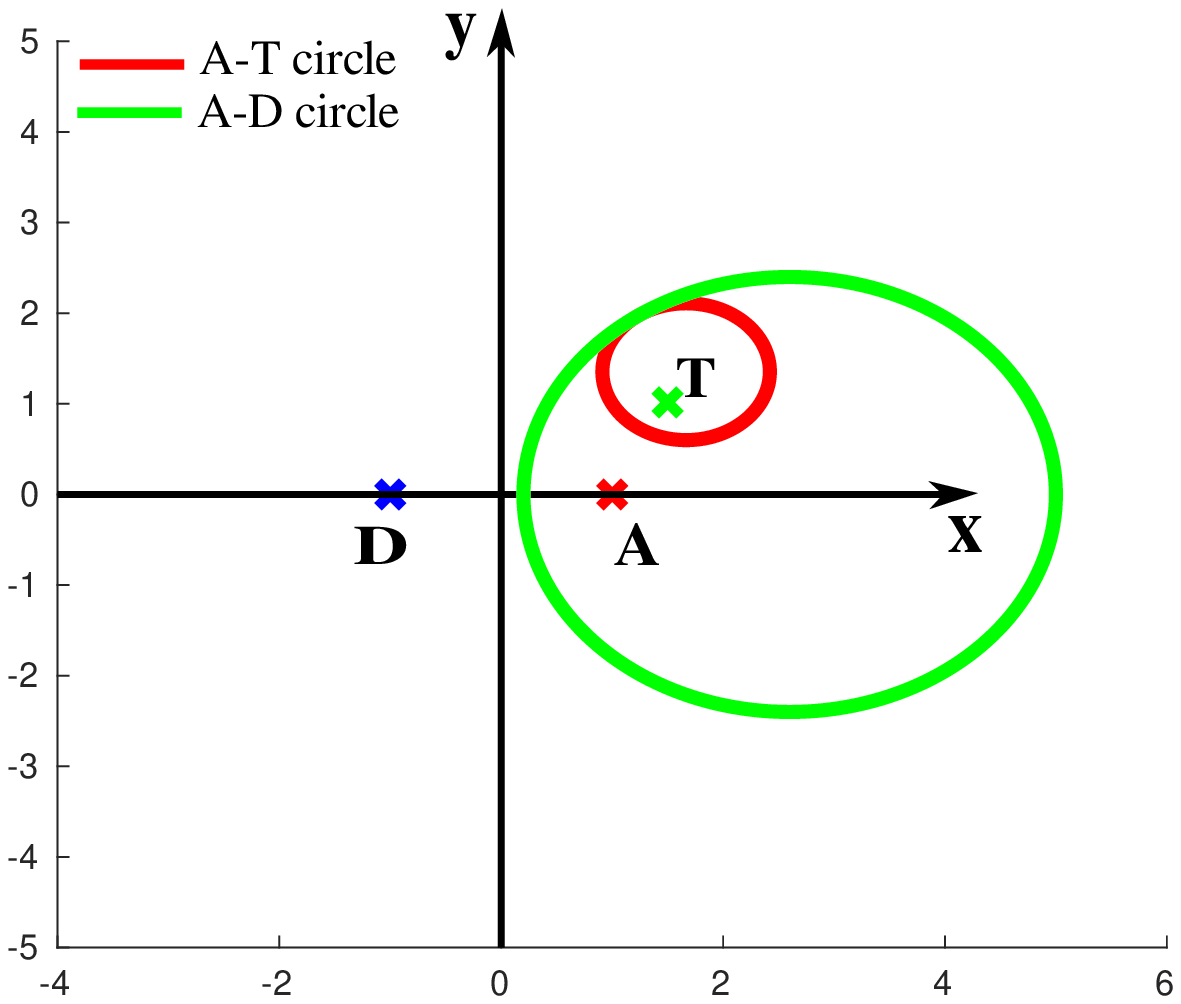}
		\caption{}
		\label{fig:gamma2_circle}
	\end{subfigure}\hfil 
	\begin{subfigure}{0.5\textwidth}
	\centering
		\includegraphics[width=6cm]{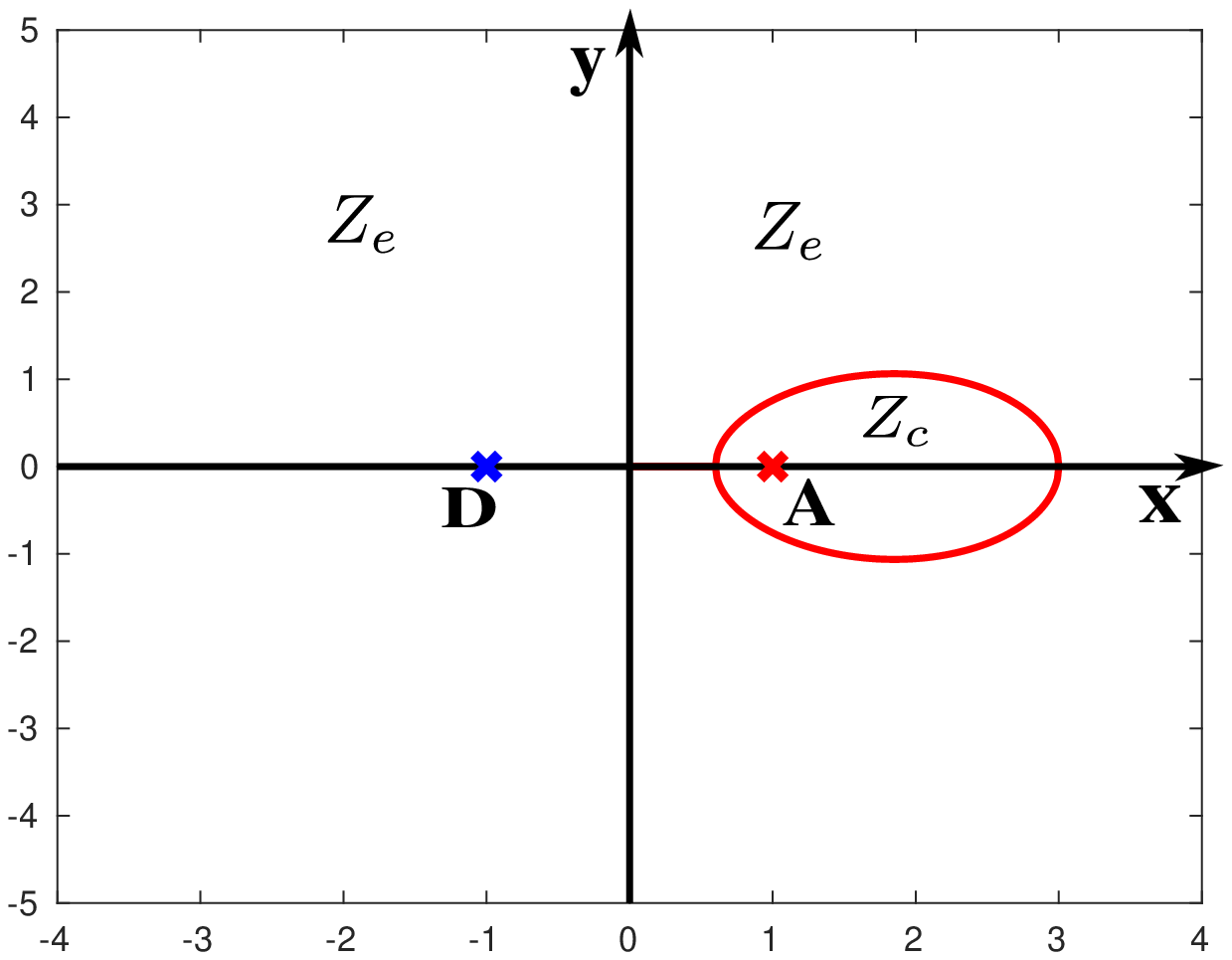}
		\caption{}
		\label{fig:gamma2_escape}
	\end{subfigure}\hfil 
	\caption{Example plots of the escape region for a constant speed target when $\gamma_{AD}>1$. (a) Apollonius circles of $A-D$ and $A-T$ engagements at the target escape boundary. (b) The curve that divides the plane into escape and capture regions for the target. }
	\label{fig:gamma2}
\end{figure*}

When $\gamma_{AD}<1$, the $A-T$ circle will be outer-tangential to the $A-D$ circle at the escape boundary of the target as shown in Fig.~\ref{fig:gamma1_circle}, and the solution to be used is given by equation~\eqref{eq:gamma_solution1}. The curve that divides the $x-y$ plane into the target escape and capture zones is shown in Fig.~\ref{fig:gamma1_escape} for an example 8x8\,m plane. 
When $\gamma_{AD}>1$, the $A-T$ circle will be inner-tangential to the $A-D$ circle at the escape boundary of the target as shown in Fig.~\ref{fig:gamma2_circle}, and the solution to be used is given by equation~\eqref{eq:gamma_solution2}. The curve that divides the $x-y$ plane into the target escape and capture zones is shown in Fig.~\ref{fig:gamma2_escape} for an example 8x8\,m plane. It can be seen that the target capture zone is reduced to a closed space beyond which the target escapes. This phenomenon is due to the speed advantage of the defender over the attacker.
\subsubsection{Stationary target case}
\begin{figure*}
	\centering 
	\begin{subfigure}{0.5\textwidth}
	\centering
		\includegraphics[width=6cm]{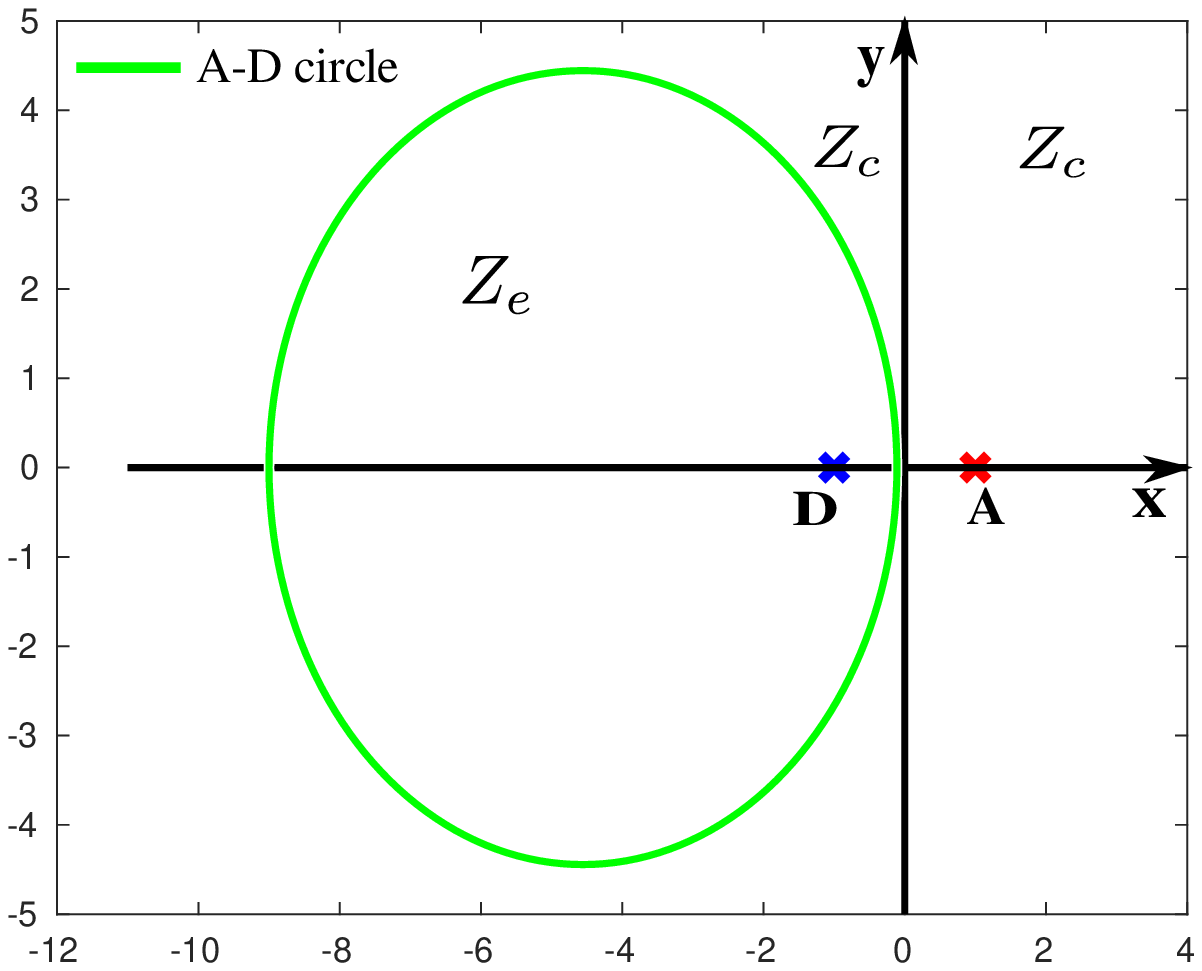}
		\caption{}
		\label{fig:stationary_gamma1_escape}
	\end{subfigure}\hfil 
	\begin{subfigure}{0.5\textwidth}
	\centering
		\includegraphics[width=6cm]{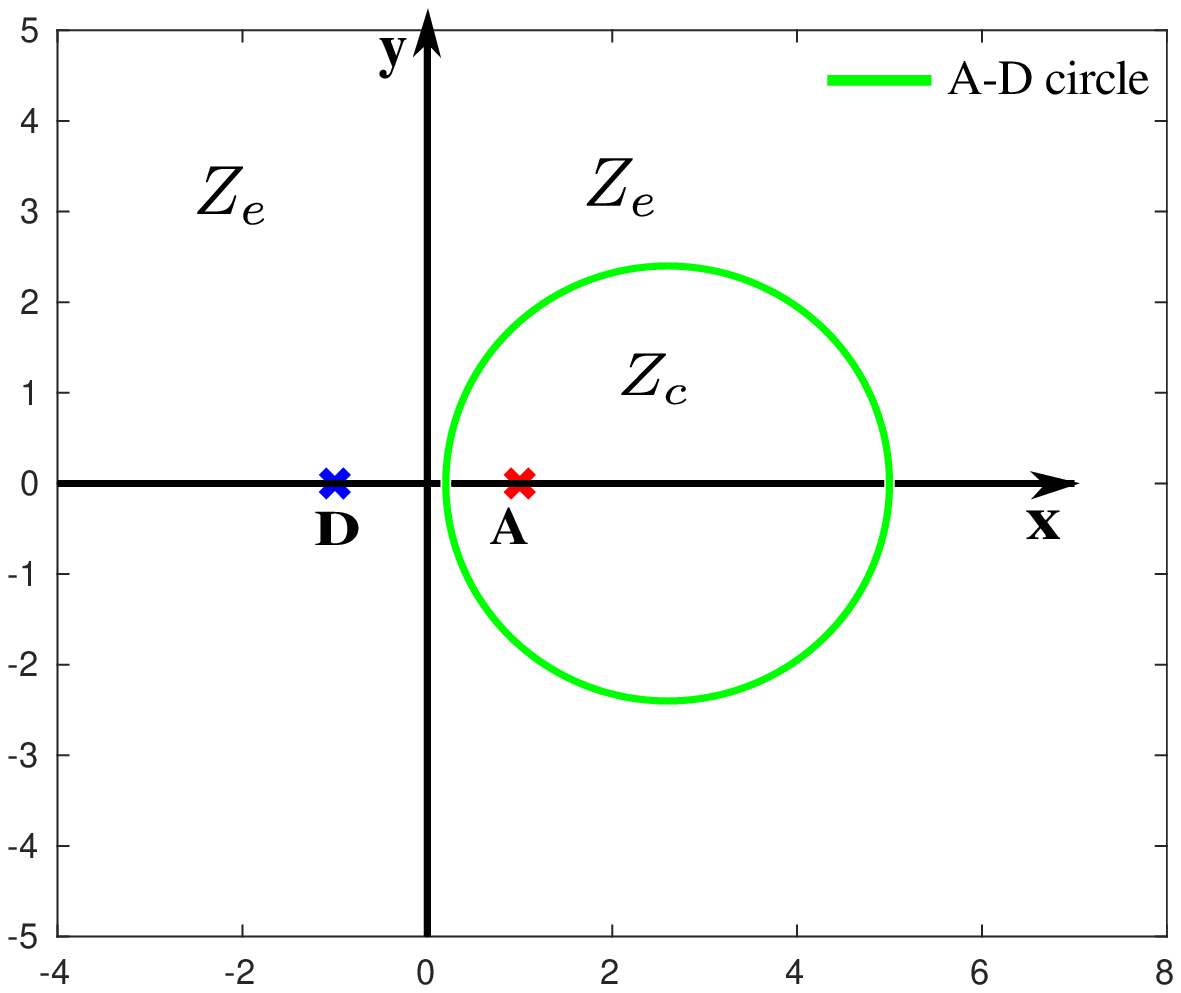}
		\caption{}
		\label{fig:stationary_gamma2_escape}
	\end{subfigure}\hfil 
	\caption{Example plots of the escape region for a stationary target when $\gamma_{AD} \neq 1$. The $A-D$ Apollonius circle divides the plane into the target escape zone, $z_e$ and capture zone, $z_c$. (a) When $\gamma_{AD}<1$ (b) when $\gamma_{AD}>1$. }
	\label{fig:stationary_gamma}
\end{figure*}

Since the $A-T$ Apollonius circle is the target point itself as given by equations \eqref{eq:stationary_target_center} and \eqref{eq:stationary_target_radius}, the curve that divides the plane into the target escape and capture zones will be the $A-D$ Apollonius circle. When $\gamma_{AD}<1$, the target escape zone will be inside the $A-D$ circle, and when $\gamma_{AD}>1$, the target escape zone will be outside the $A-D$ circle. An example for both cases is shown in Fig.~\ref{fig:stationary_gamma}.
\subsubsection{Variable target velocity case}
\begin{figure*}
	\centering 
	\begin{subfigure}{0.5\textwidth}
	\centering
		\includegraphics[width=6cm]{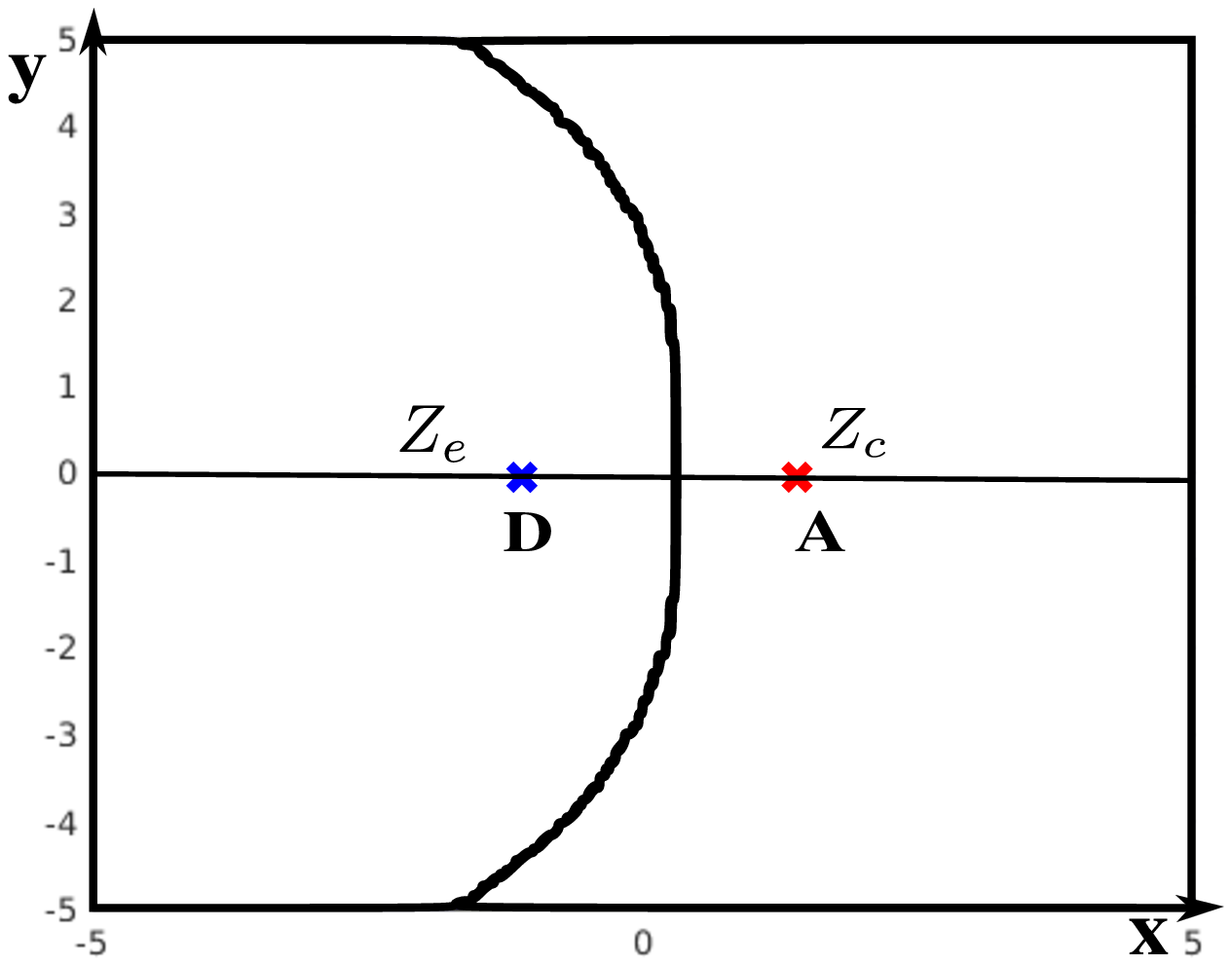}
		\caption{}
		\label{fig:variable_gamma1}
	\end{subfigure}\hfil 
	\begin{subfigure}{0.5\textwidth}
	\centering
		\includegraphics[width=6cm]{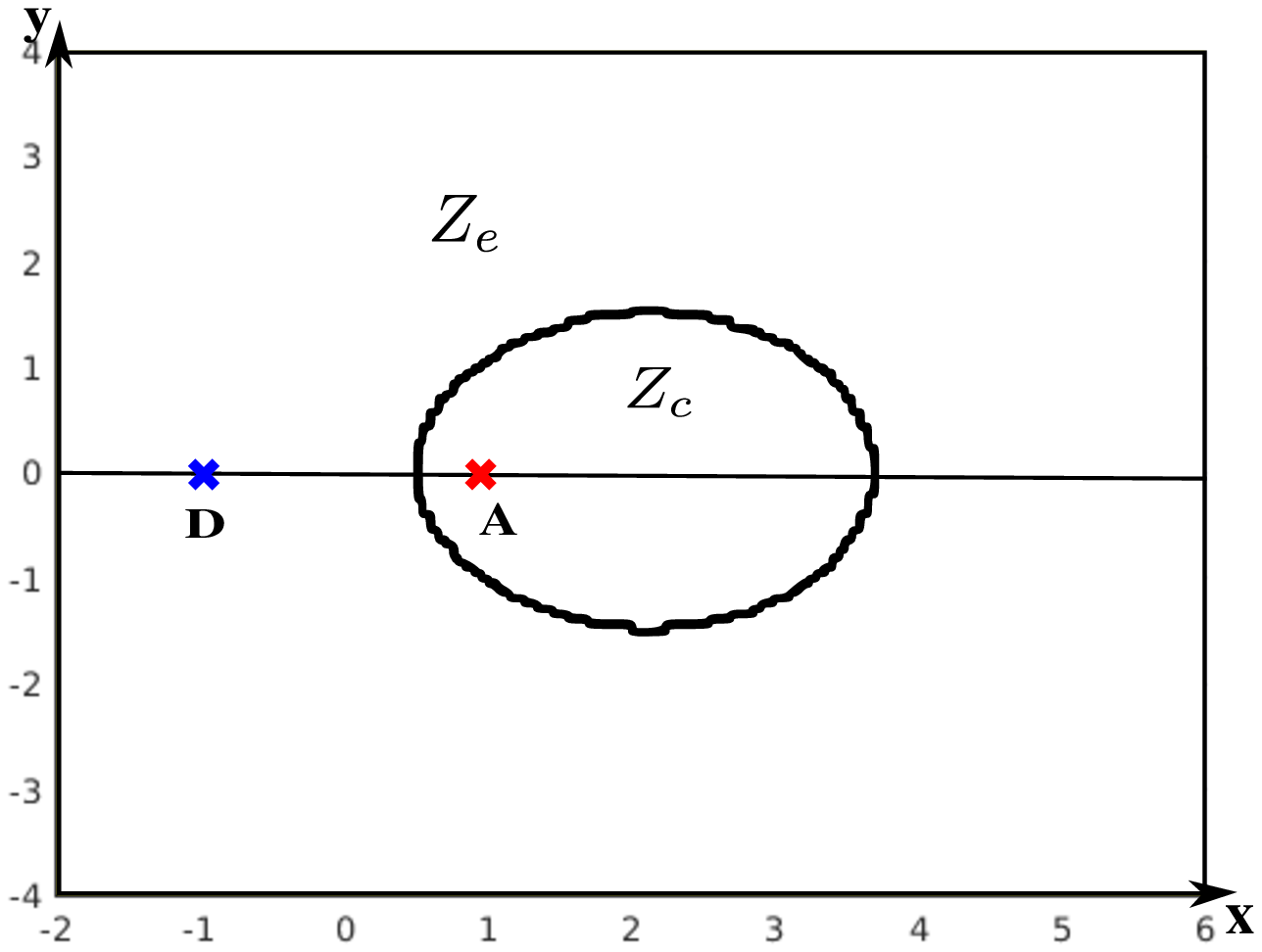}
		\caption{}
		\label{fig:variable_gamma2}
	\end{subfigure}\hfil 
	\caption{Example plots of the escape region for a variable velocity target when $\gamma_{AD} \neq 1$. (a) When $\gamma_{AD}<1$ (b) when $\gamma_{AD}>1$. }
	\label{fig:variable_gamma}
\end{figure*}

This case needs to be numerically computed similar to the analysis given in Sec.~\ref{sec:var_target_escape}. The only difference here is that instead of checking if the $A-T$ Apollonius circle crosses the $y-$axis, we need to check if the $A-T$ circle crosses the $A-D$ circle. An example for the curve that divides the $x-y$ plane into the target escape and capture zones is shown in Fig.~\ref{fig:variable_gamma}. 
\subsection{Escape region based on the estimated states (stochastic escape region)}
\begin{figure}
	\centering
	\includegraphics[width=6cm]{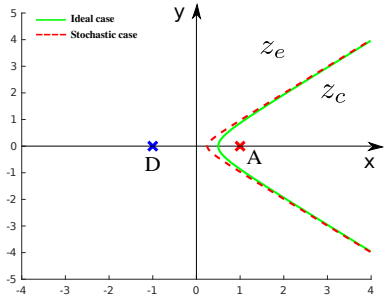}
	\caption{Stochastic escape zone for the constant speed target, $T$. $Z_e$ is the escape zone and $Z_c$ is the capture zone. }
	\label{fig:sigma_region}
\end{figure}

The analysis presented in Sec.~\ref{sec:CS_target_escape} is based on ideal conditions with a deterministic system, and the true position values of the agents are known. Nevertheless, the analysis can easily be converted to include stochasticity and attacker state estimation. The EKF formulated in Sec.~\ref{sec:ekf} allows the attacker states to be estimated precisely within the $\pm3\sigma$ bounds. If the worst case (for the target) of $-3\sigma$ is considered, the dividing curve shifts a little to the left, reducing the escape zone of the target as shown in Fig.~\ref{fig:sigma_region}. A similar analysis can be performed on all the other target speed cases given in the paper, and the corresponding results are not included here due to space constraints.

\section{Simulation results} \label{sec:results}
The efficacy of the proposed NMPC scheme is evaluated through extensive numerical simulations. The simulations were performed using MATLAB R2017b and VirtualArena \cite{c17} on an Ubuntu 16.04, Intel i5 PC with 8GB RAM running at 2.4GHz. Initially, we present the results for the constant speed target case followed by the variable target velocity case.
\subsection{Simulation setup}
The look-ahead window for the NMPC is $6$ steps with a sampling time of $0.05$\,s, which makes the prediction window $ \tau_{h} =0.3$\,s for all simulations. The covariance matrices $ Q $ and $ \Sigma $ for the EKF are selected as
\begin{equation*}
	Q=\begin{bmatrix}
	0.1 & 0 & 0 & 0\\
	0&0.1&0 & 0\\
	0&0&0.01 & 0\\
	0 & 0 & 0 & 0.1
	\end{bmatrix},
	\Sigma=\begin{bmatrix}
	0.1 & 0 & 0&0\\
	0&0.1&0&0\\
	0&0&0.01&0\\
	0&0&0&0.01
	\end{bmatrix}.
\end{equation*} 
For the simulation purpose, the attacker switches the guidance law between pure pursuit (PP) and proportional navigation (PN) at each second. This technique is opted in order to show that the attacker is intelligent and also the efficacy of the proposed NMPC scheme against unknown attacker guidance laws. Note that the attacker guidance law is not known to the target-defender team. The PP guidance law is given as \cite{siouris2004missile}
$	a_{A}=-\kappa\left( \alpha_{A}-\theta\right), $
and the PN guidance is given as \cite{c13}    
$	a_{A}=Nv_{A}\dot{\theta}$. 
The navigation constant, $N$ = 3, is taken for the PN guidance law and $\kappa$ = 2 for the PP law. $ R_{c}=1$\,m and $ r_{c}=1 $\,m are the attacker-target and the attacker-defender capture radii. The angular velocities of the agents are constrained to $-0.5\leq \{\dot{\alpha}_{T},\dot{\alpha}_{D}\} \leq 0.5$\,rad/s considering the real-world implementation constraints. The maximum velocity of the target is also constrained to $\bar{v}_T \leq 2$\,m/s to keep the $A-T$ speed ratio $\gamma_{AT} < 1$. Initial conditions of the agents for each simulation is given in Table~\ref{table:initial}, \ref{table:initial_unequal}, and are represented in Fig.~\ref{fig:cs_map},\ref{fig:var_map}, and \ref{fig:variable_gamma2_map}.	

\begin{table*}
	\centering
	\begin{tabular}{ |c|c|c|c|c|c|c|c|c|c|c| } 
		\hline
		\multicolumn{3}{|c|}{constant speed target}&
		\multicolumn{4}{|c|}{variable velocity target} & \multicolumn{4}{|c|}{Comparison}\\
		\hline 
		parameter&escape&capture&parameter&escape&escape&capture & parameter & NMPC&A-CLOS &CLOS\\
		&&&&$e$ not violated& $e$ violated&&&&&\\
		\hline
		$ (x_{A},y_{A}) $ (m) & (50,50) & (50,50) & $(x_{A},y_{A}) $ (m) & (50,0) & (50,0)& (50,0) &$ (x_{A},y_{A}) $ (m) & (0,5000) & (0,5000)& (0,5000) \\
		$ (x_{T},y_{T}) $ (m) & (25,30) & (60,130) & $(x_{T},y_{T}) $ (m) & (-50,100) & (10,100)& (100,100) &$ (x_{T},y_{T}) $ (m) & (0,0) & (0,0)& (0,0) \\
		$ (x_{D},y_{D}) $ (m) & (0,0) & (0,0) & $(x_{D},y_{D}) $ (m) & (-50,0) & (-50,0) & (-50,0) &$ (x_{D},y_{D}) $ (m) & (0,0) & (0,0)& (0,0) \\
		$ \alpha_{A} $ (rad)  &  -2.2   &  0.78 & $ \alpha_{A} $ (rad) &  1.57   & 3.14 &  1.57 &$ \alpha_{A} $ (rad)  &  -1.57   &  -1.57&  -1.57   \\
		$ \alpha_{T} $ (rad)  &  -2.2   &  0.78 & $u_{x} $ (m/s)  &  0 & 0 &  0   &$ \alpha_{T} $ (rad)  &  0   &  0&  0   \\
		$ \alpha_{D} $ (rad)  &  0.78   &  0.78 & $u_{y} $ (m/s)  &  0 & 0 &  0   &$ \alpha_{D} $ (rad)  &  1.57   &  1.57&  1.57   \\
		$ v_{A} $ (m/s)  &  4   &   4 & $\alpha_{D} $ (rad)  &  0.78 & 0 &   0 &$ v_{A} $ (m/s)  &  600   &   600&   600  \\
		$ v_{T} $ (m/s)  &  2   &   2 & $v_{A} $ (m/s)  &  4   & 4 & 4  &$ v_{T} $ (m/s)  &  200   &   200 &   200  \\
		$ v_{D} $ (m/s)  &  4   &   4   &$v_{D}$ (m/s) & 4 & 4 & 4 &$ v_{D} $ (m/s)  &  600   &   600&   600  \\
		\hline
	\end{tabular}
	\caption{Initial parameters for the agents for escape and capture scenarios.}
	\label{table:initial}
\end{table*}

\subsection{Constant speed target}
\begin{figure}
	\centering
	\includegraphics[width=6cm]{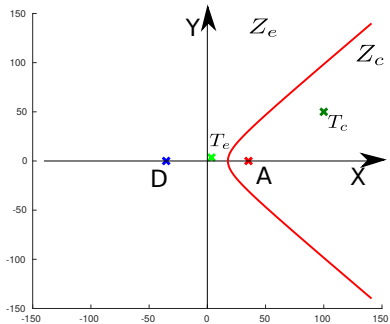}
	\caption{Initial agent configurations for the constant speed target case. }
	\label{fig:cs_map}
\end{figure}

Given the initial engagement geometry, we can say whether the target will be captured or not using the analysis given in Sec.~\ref{sec:CS_target_escape}. We validate this claim through simulation for the initial configuration given in Fig.~\ref{fig:cs_map}. The initial positions of the attacker and the defender are selected as $A(35,0)$ and $D(-35,0)$. The $A-D$ speed ratio, $\gamma_{AD}=1$, and the $A-T$ speed ratio, $\gamma_{AT}=0.5$. 

\subsubsection{Target escape case}
\begin{figure*}
\centering
	\begin{subfigure}{6cm}
		\includegraphics[width=6cm]{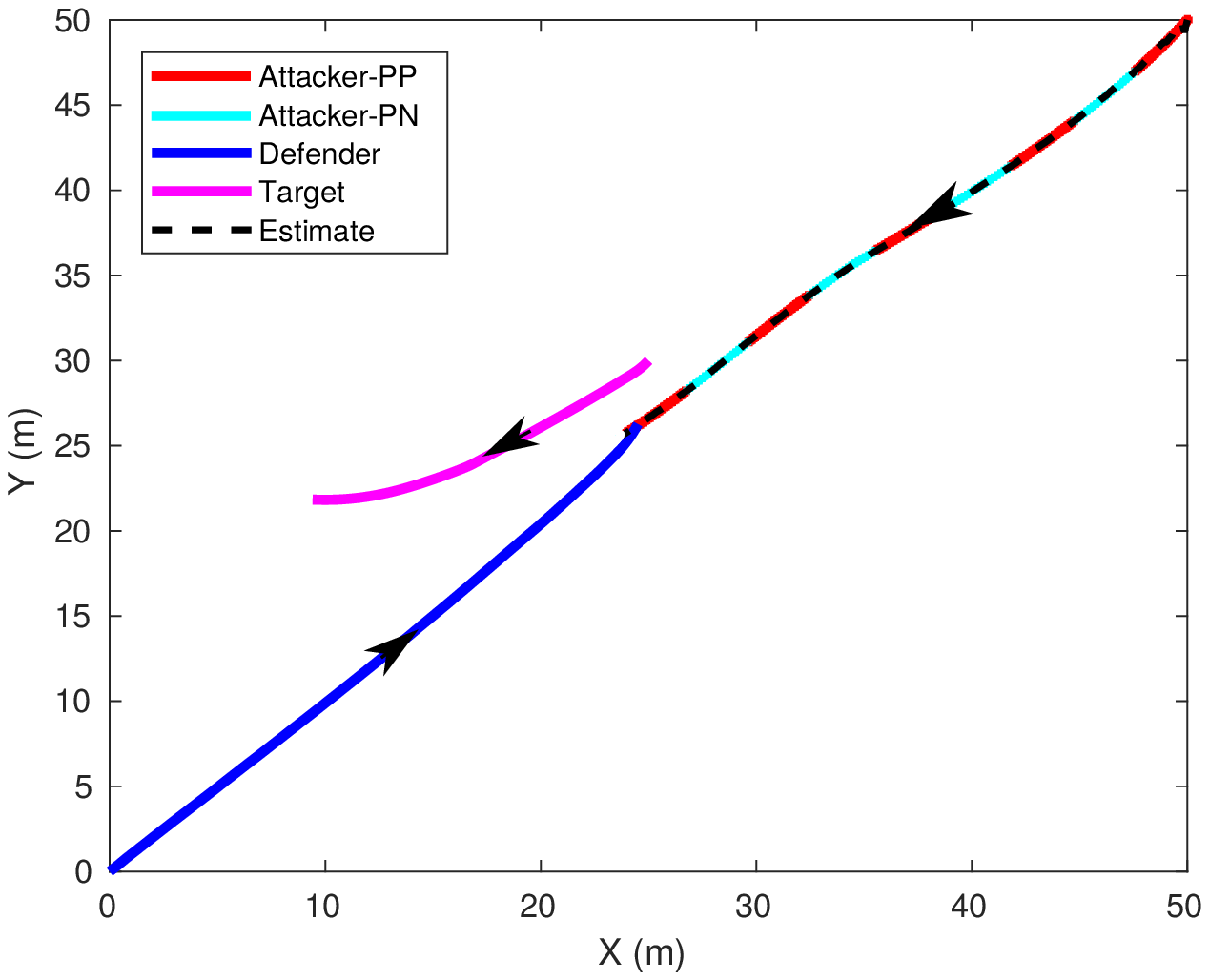}
		\caption{}
		\label{fig:cs_escape_traj}
	\end{subfigure}\hspace{1cm}
	\begin{subfigure}{6cm}
		\includegraphics[width=6cm]{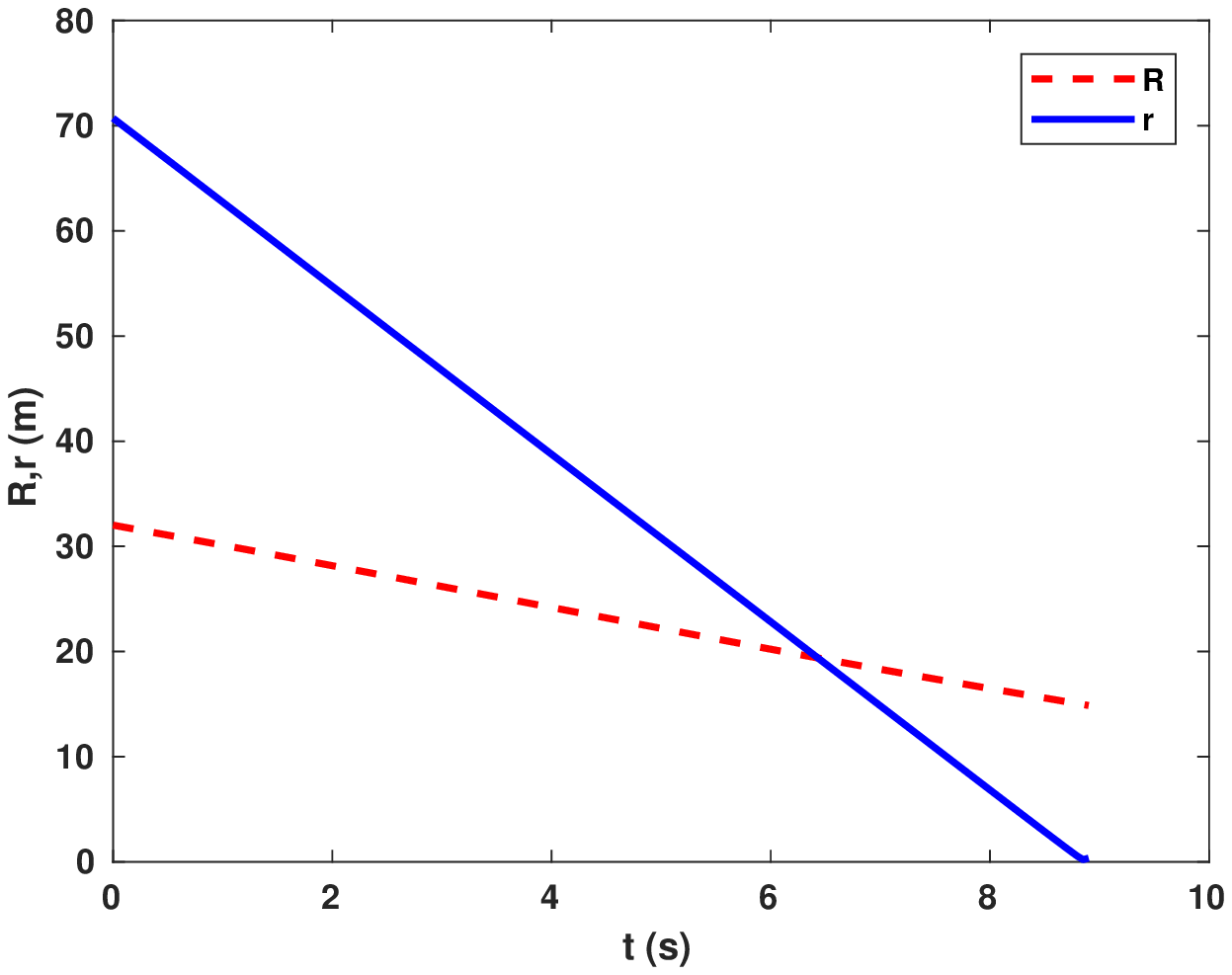}
		\caption{}
		\label{fig:cs_escape_r_R}
	\end{subfigure}
    
	\begin{subfigure}{6cm}
		\includegraphics[width=6cm]{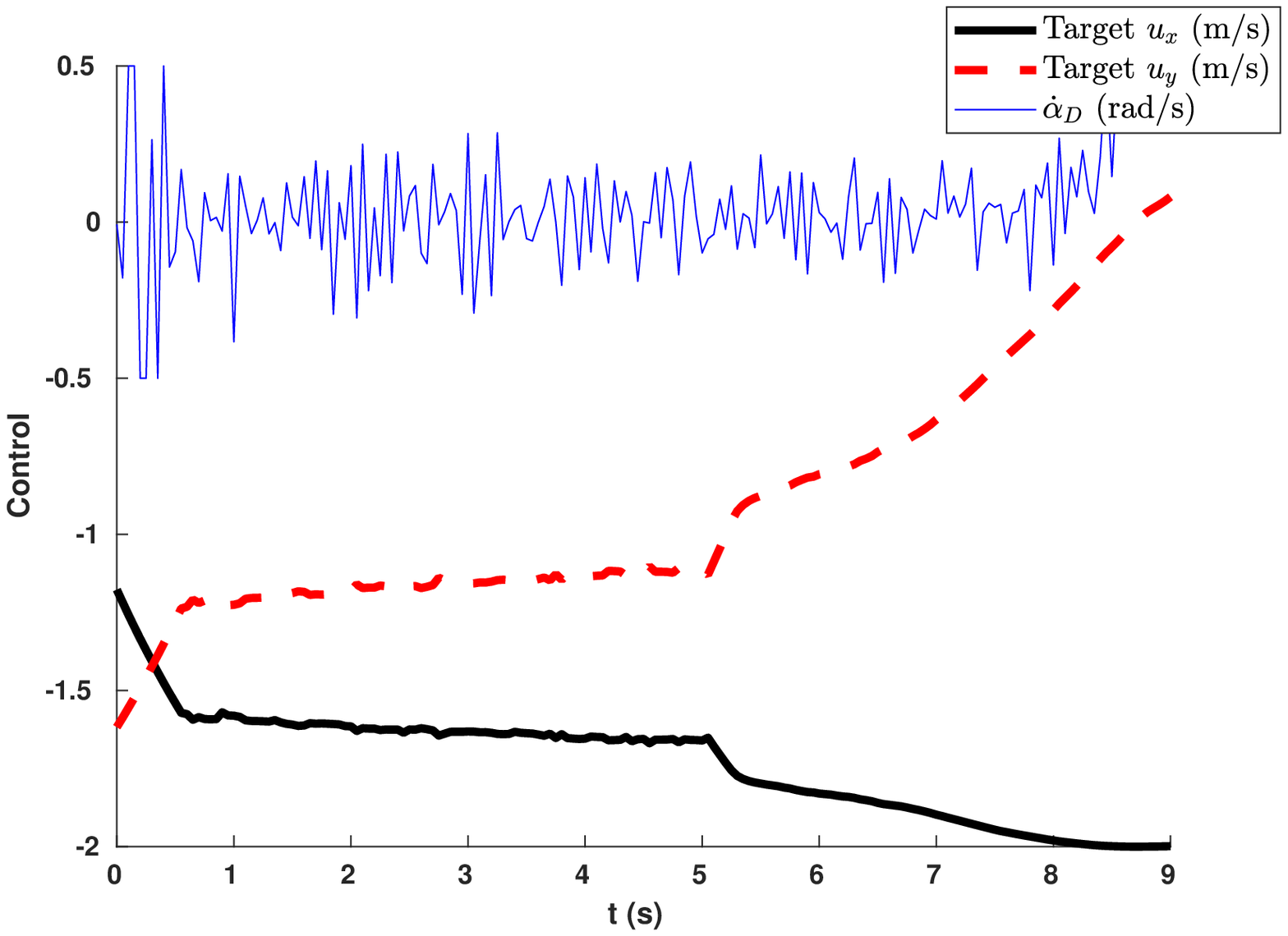}
		\caption{}
		\label{fig:cs_escape_control}
	\end{subfigure}\hspace{1cm}
	\begin{subfigure}{6cm}
		\includegraphics[width=6cm]{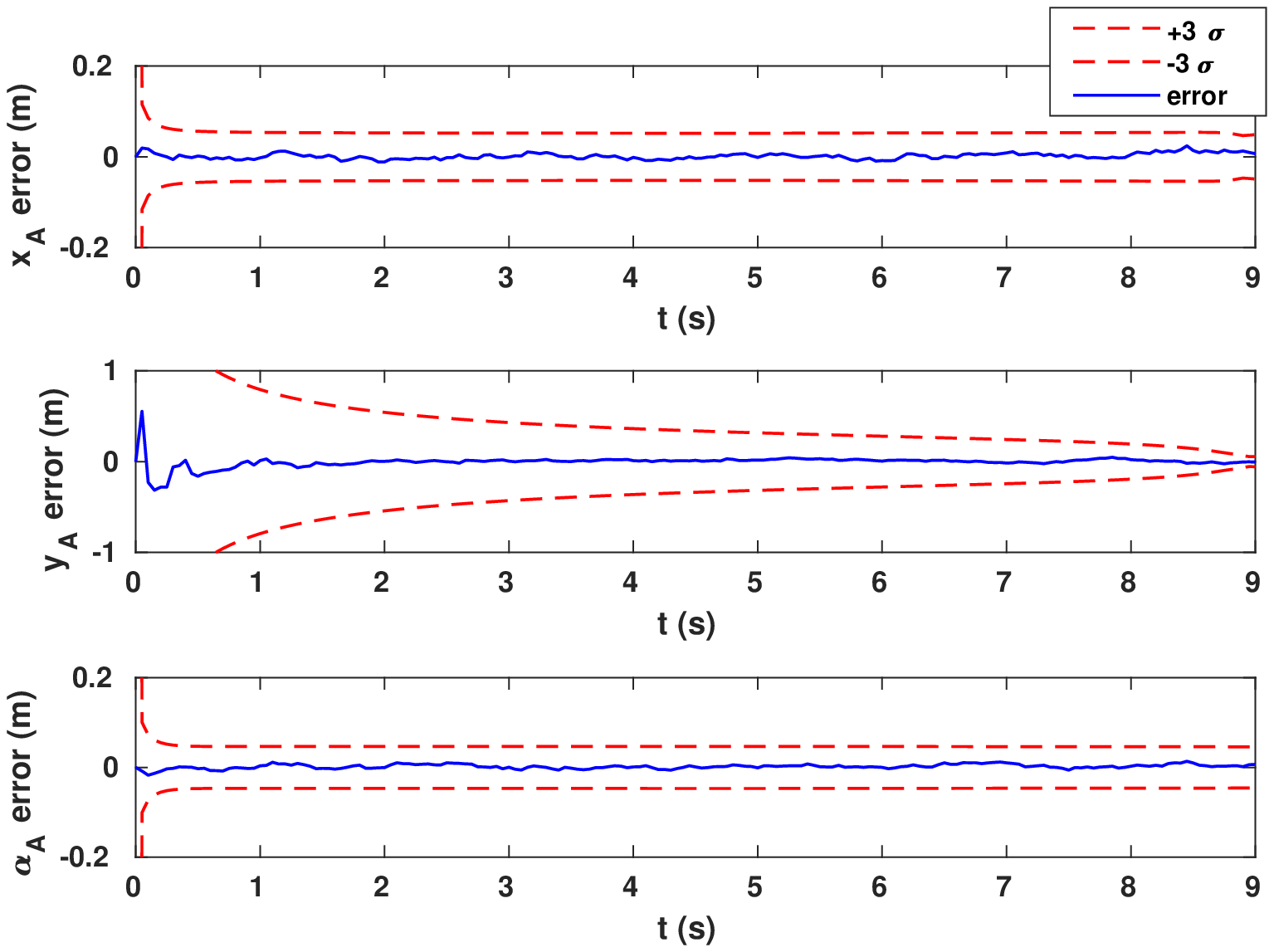}
		\caption{}
		\label{fig:cs_escape_error}
	\end{subfigure}
		\caption{Target escape scenario for the constant speed target. (a) Trajectories of the agents. (b) Distance between the agents. (c) Optimal control inputs determined by the NMPC. (d) Error in the attacker state estimates.}
		\label{fig:cs_escape}
\end{figure*}

The initial position of the target is selected inside the escape zone $Z_e$ and is represented by $T_e(3.5,3.5)$ in Fig.~\ref{fig:cs_map}. The initial conditions for the simulation is given in Table~\ref{table:initial} escape column of constant speed case. The agent trajectories are shown in Fig.~\ref{fig:cs_escape_traj}, and the evolution of the distances ($R$ and $r$) between the agents are shown in Fig.~\ref{fig:cs_escape_r_R}.
Since the attacker guidance law is changing over time, we represent this change by using different color sequence in the plot, PP in red and PN in cyan. Even with the attacker guidance law switching from PP and PN, the defender was able to intercept the attacker and mitigate the target capture. The optimal control inputs determined by the NMPC are given in Fig.~\ref{fig:cs_escape_control}. It can be seen that the bounds on the angular velocity of the defender were strictly followed. Fig.~\ref{fig:cs_escape_error} shows the errors in the estimates of the attacker position and heading angle. The errors are low and stay within the $ 3\sigma $ bounds, which implies that the estimator performance is satisfactory. 

\subsubsection{Target capture case}
\begin{figure*}%
    \centering
    \subfloat[\centering \label{fig:cs_capture_a}]{{\includegraphics[width=6cm]{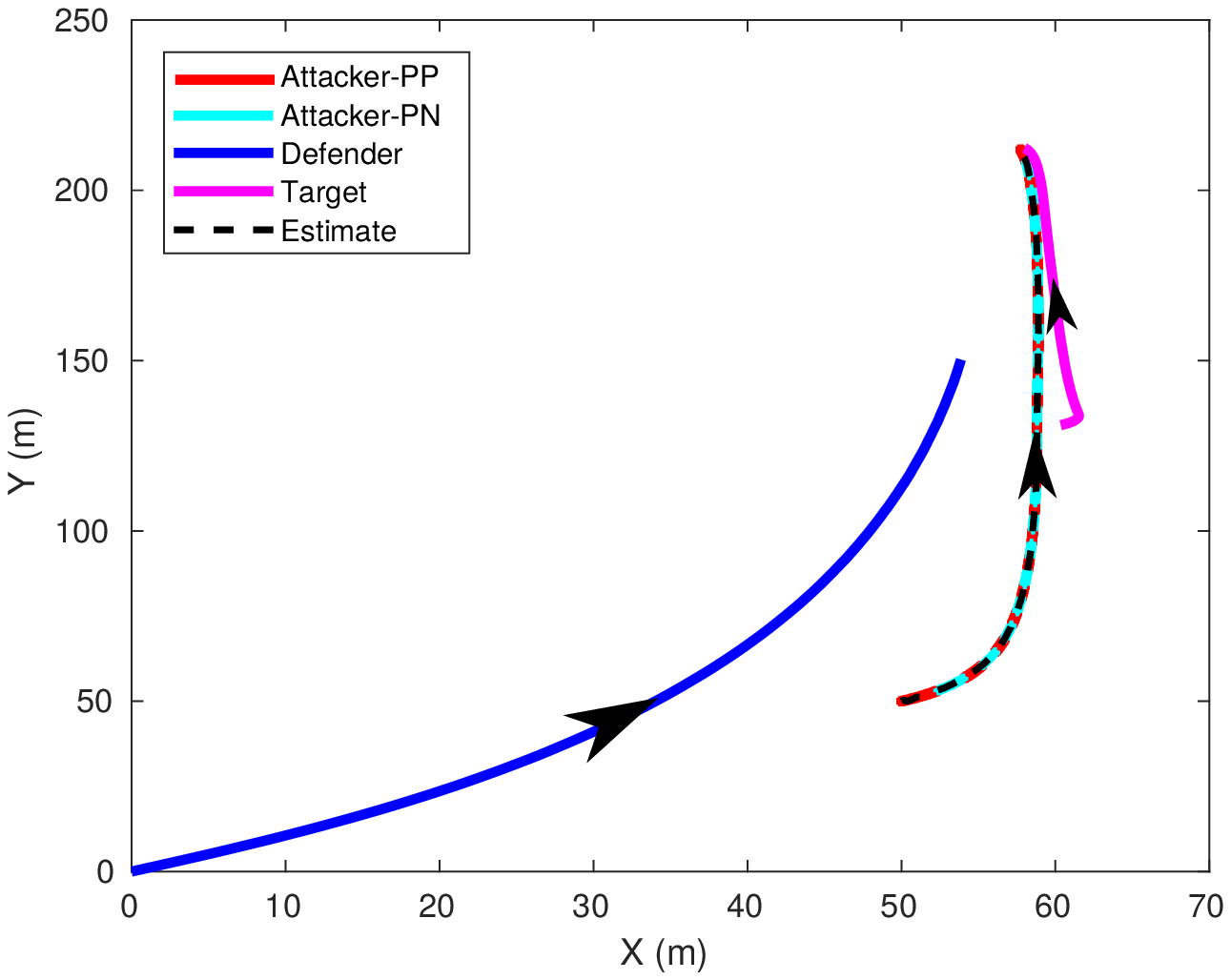} }}%
    \subfloat[\centering \label{fig:cs_capture_b}]{{\includegraphics[width=6cm]{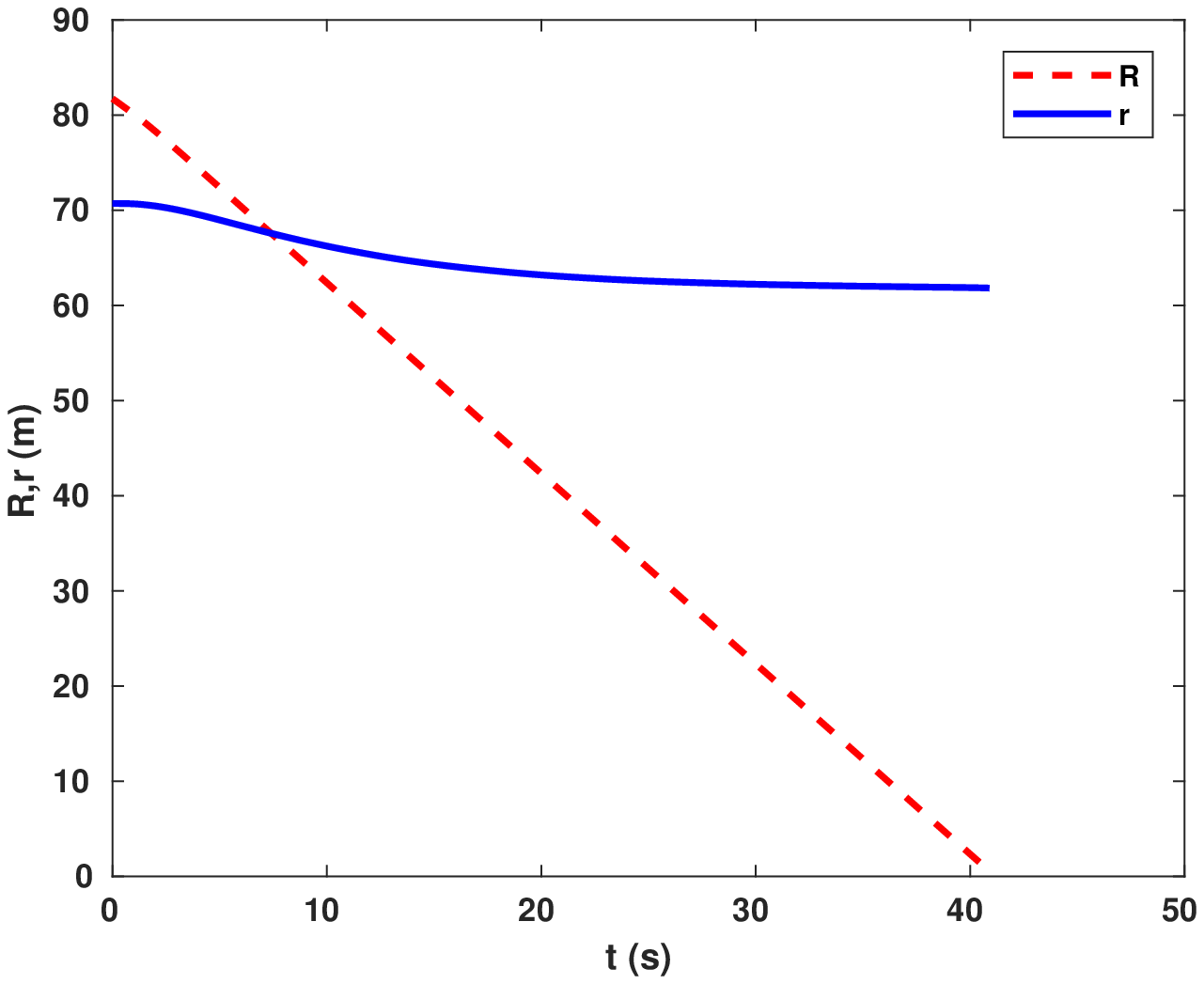} }}\\
    \subfloat[\centering \label{fig:cs_capture_control}]{{\includegraphics[width=6cm]{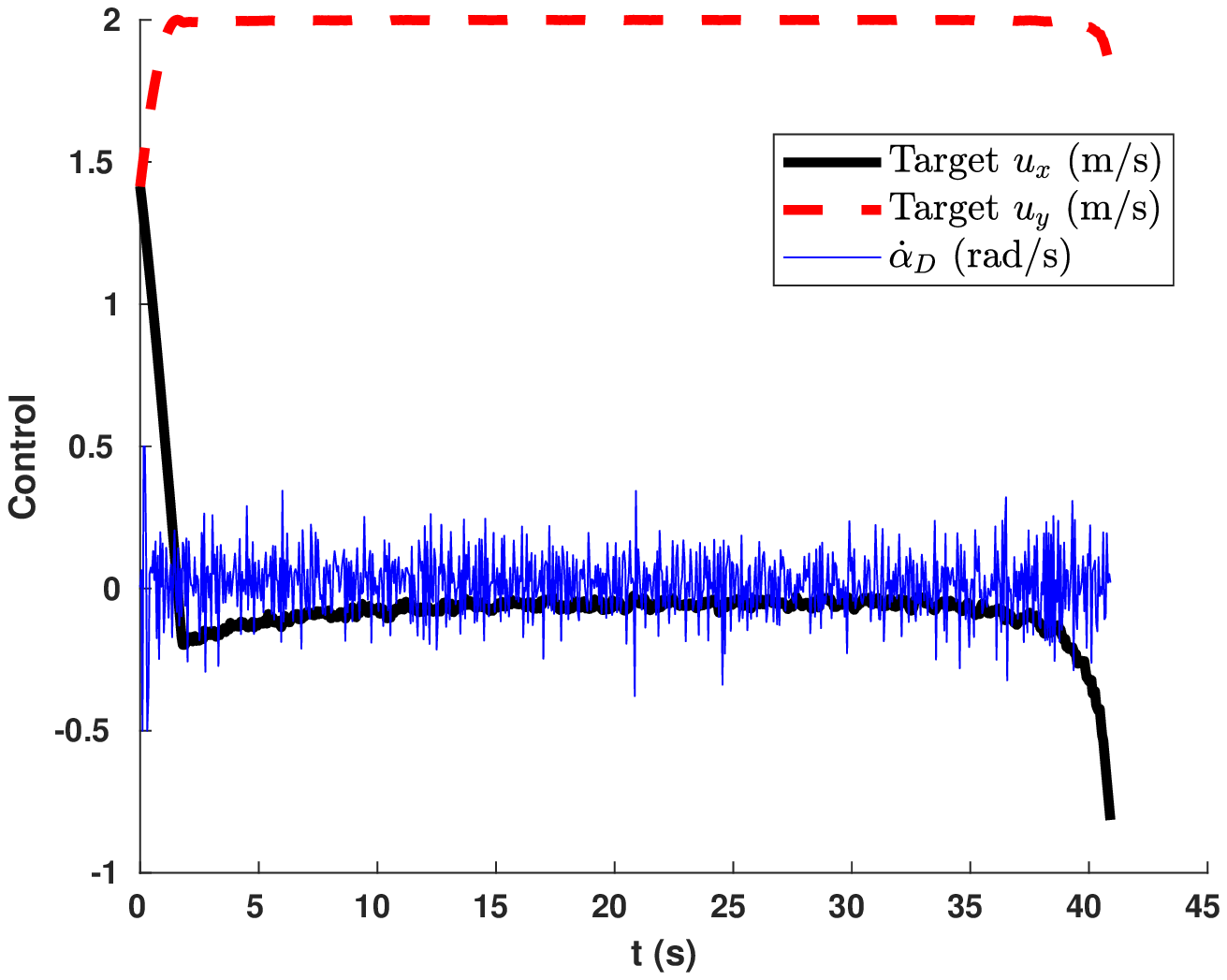} }}%
    \subfloat[\centering \label{fig:cs_capture_error}]{{\includegraphics[width=6cm]{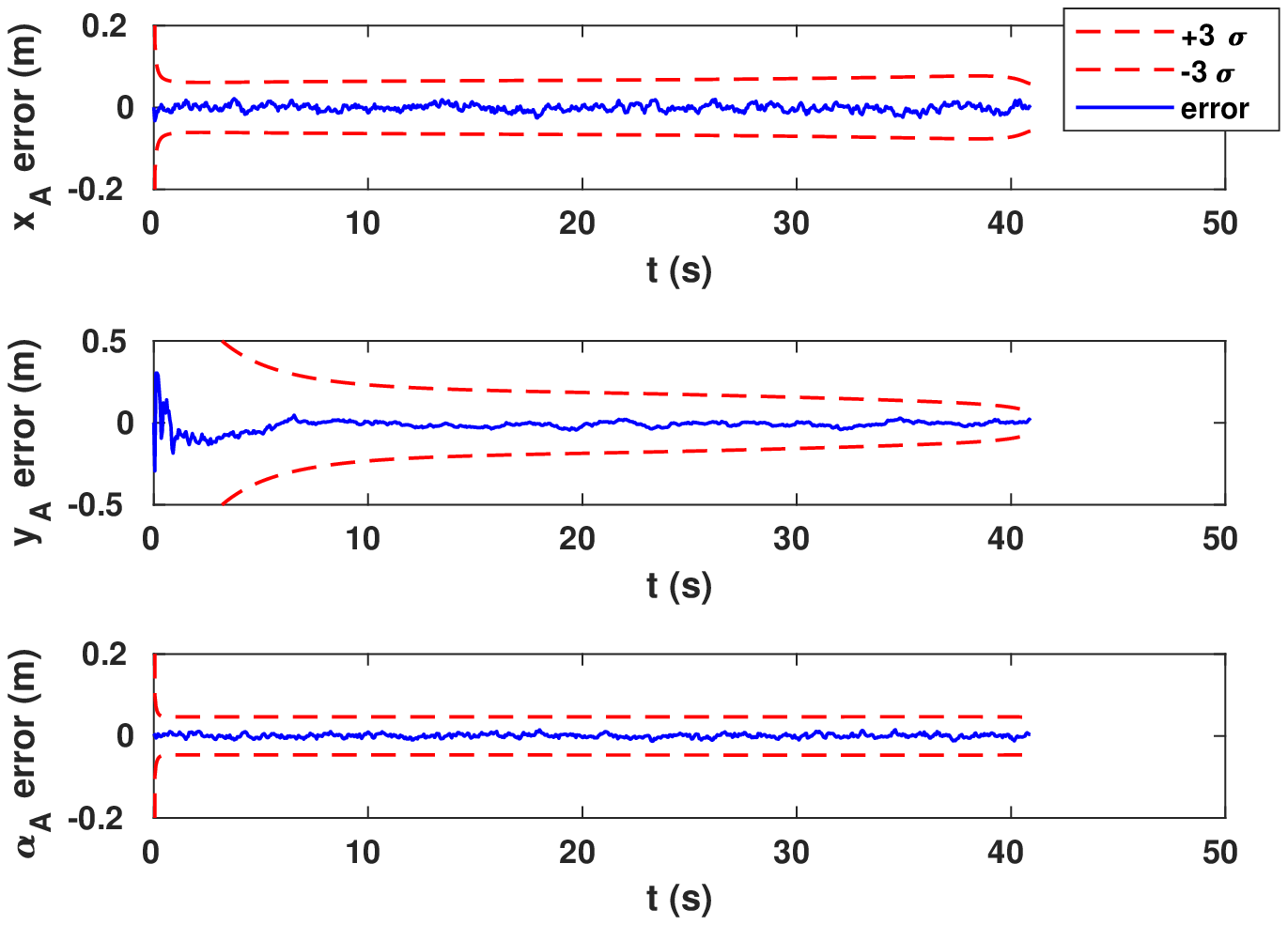} }}%
    \caption{Target capture scenario for the constant speed target. (a) Trajectories of the agents. (b) Distance between the agents. (c) Optimal control inputs determined by the NMPC. (d) Error in the attacker state estimates.}%
    \label{fig:cs_capture}%
\end{figure*}

The initial position of the target is selected inside $Z_c$ and is represented by $T_c(100,50)$ in Fig.~\ref{fig:cs_map}. The simulation parameters are shown in Table~\ref{table:initial} capture column of constant speed target. The agent trajectories and the evolution of the distances for this scenario are shown in Fig.~\ref{fig:cs_capture_a} and Fig.~\ref{fig:cs_capture_b}. The attacker captured the target before the defender could intercept the attacker. It can be seen from Fig.~\ref{fig:cs_capture_b} that the distance between the attacker and the defender remained almost constant since they have equal velocities and are on a tail-chase engagement. The optimal control inputs and the error in attacker states are given in Fig.~\ref{fig:cs_capture_control} and Fig.~\ref{fig:cs_capture_error} respectively. The target velocity stays constant at the specified value of 2\,m/s throughout the engagement. The estimator performance was satisfactory since the errors never crossed the upper and lower covariance bounds. 

\subsection{Variable target velocity case}
\begin{figure}
	\centering
	\includegraphics[width=6cm]{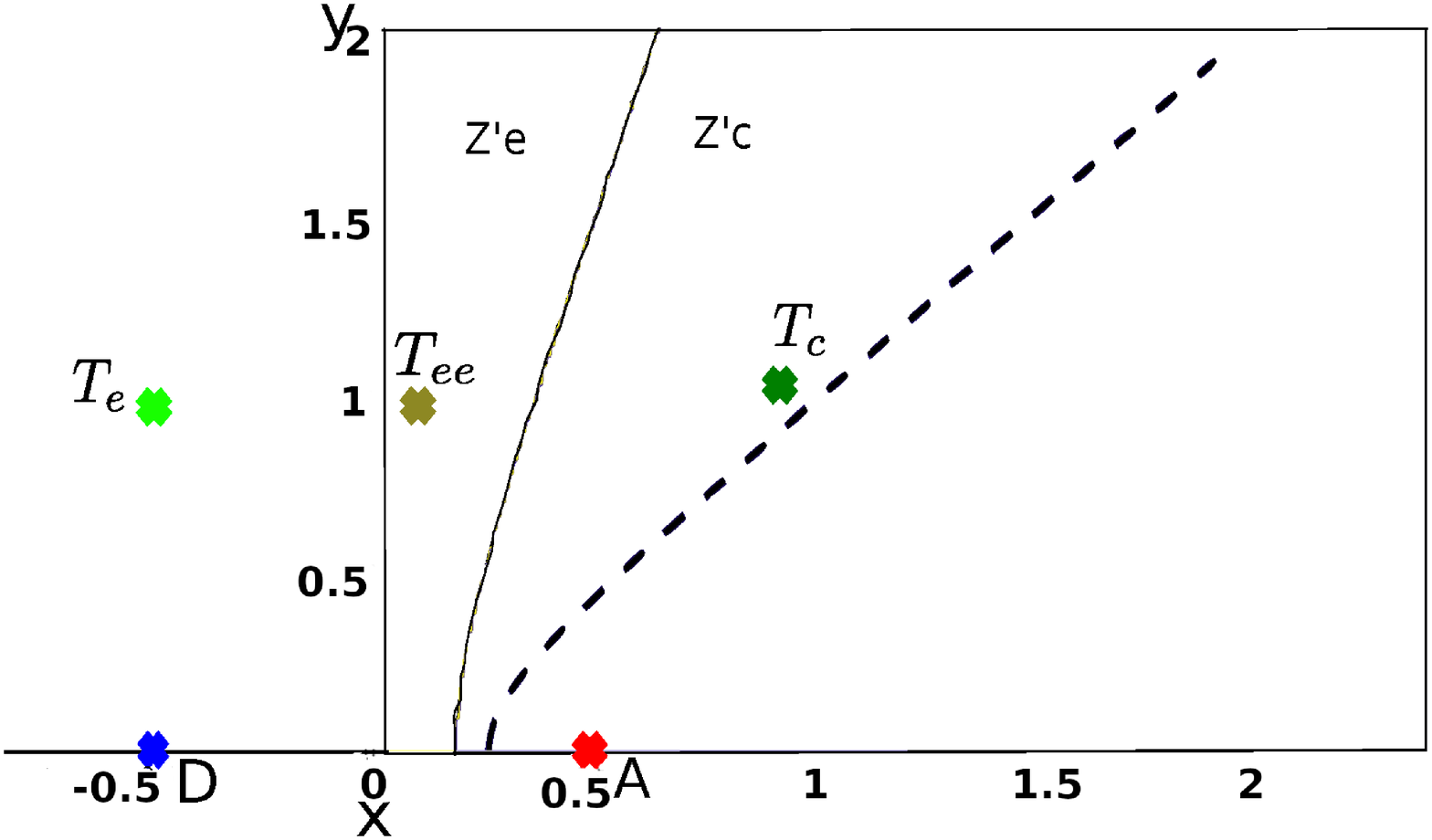}
	\caption{Initial agent configurations for the variable velocity target case. The dashed curve represents the constant speed case, the solid curve the variable velocity case, and the $y-$axis represents the boundary for the stationary target. }
	\label{fig:var_map}
\end{figure}

The analysis given for variable velocity target in Sec.~\ref{sec:var_target_escape} is validated using the initial configurations given in Fig.~\ref{fig:var_map}. The initial positions of the attacker and the defender are selected as $A(0.5,0)$ and $D(-0.5,0)$. The $A-D$ speed ratio, $\gamma_{AD}=1$, and the $A-T$ speed ratio, $\gamma_{AT}=0.5$. The safe distance parameter $e$ is selected as half of the initial distance between the target and the attacker.

\subsubsection{Target escape case where safe distance is not violated}
\begin{figure*}
	\centering 
	\begin{subfigure}{6cm}
		\includegraphics[width=\linewidth]{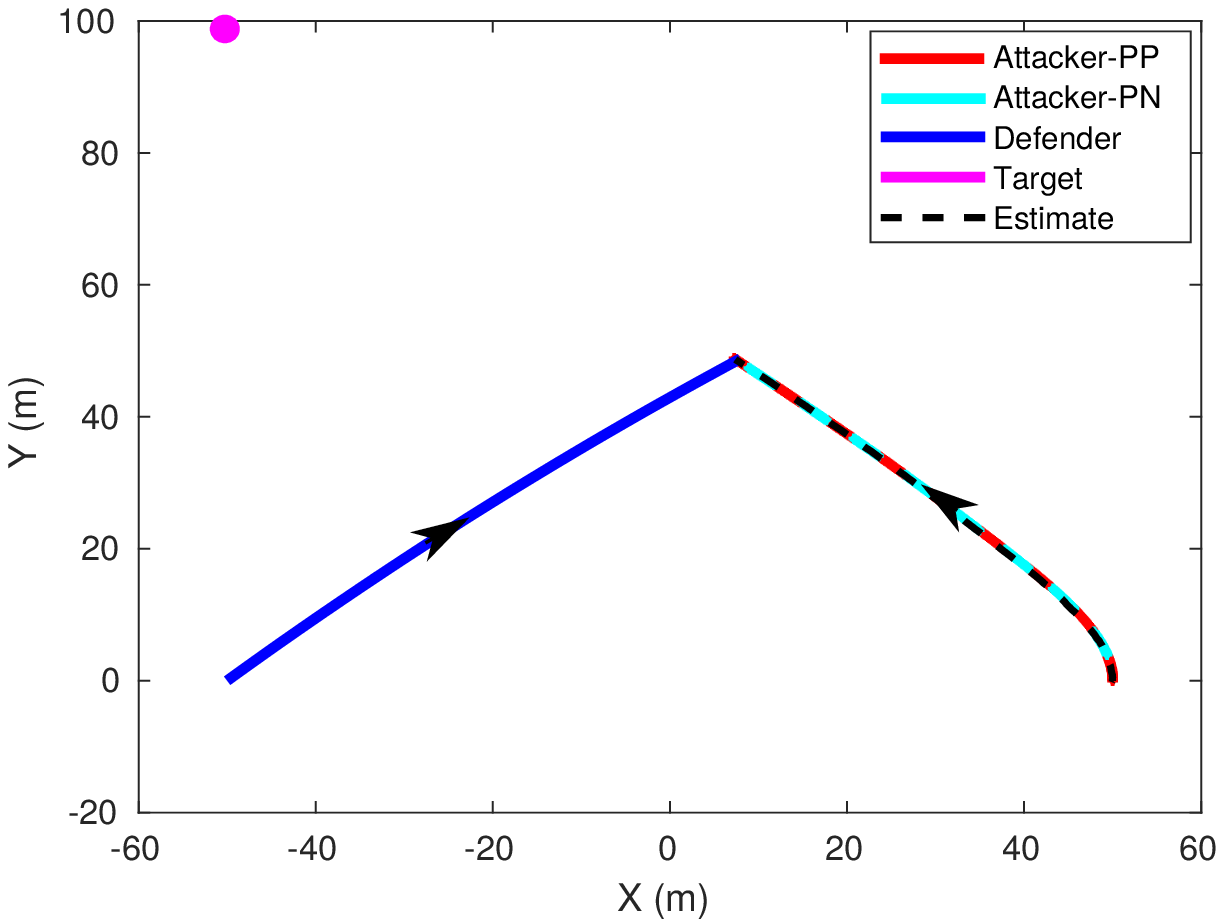}
		\caption{}
		\label{fig:var_escape_traj}
	\end{subfigure}\
	\begin{subfigure}{6cm}
		\includegraphics[width=\linewidth]{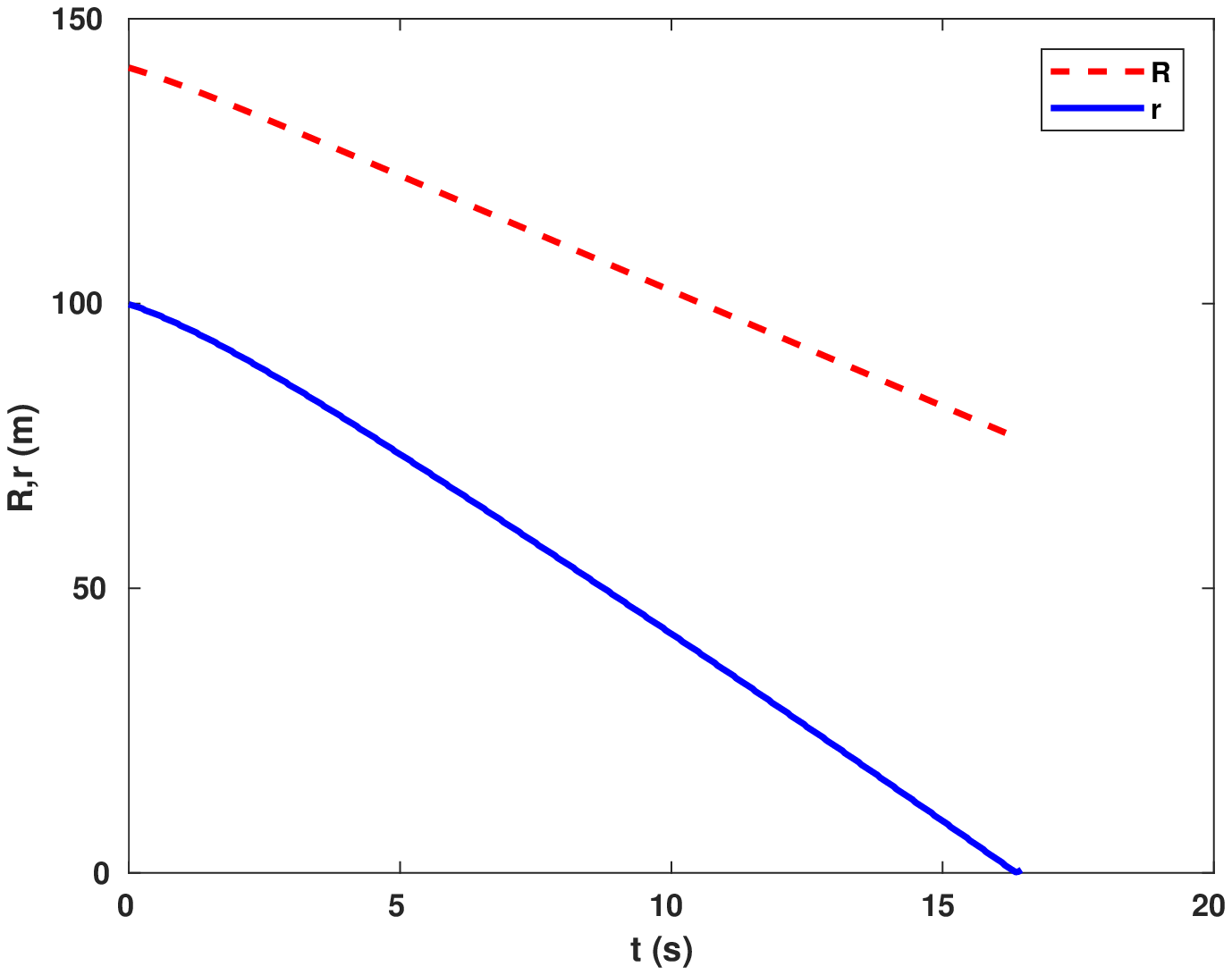}
		\caption{}
		\label{fig:var_escape_r_R}
	\end{subfigure}
    
	\begin{subfigure}{6cm}
		\includegraphics[width=\linewidth]{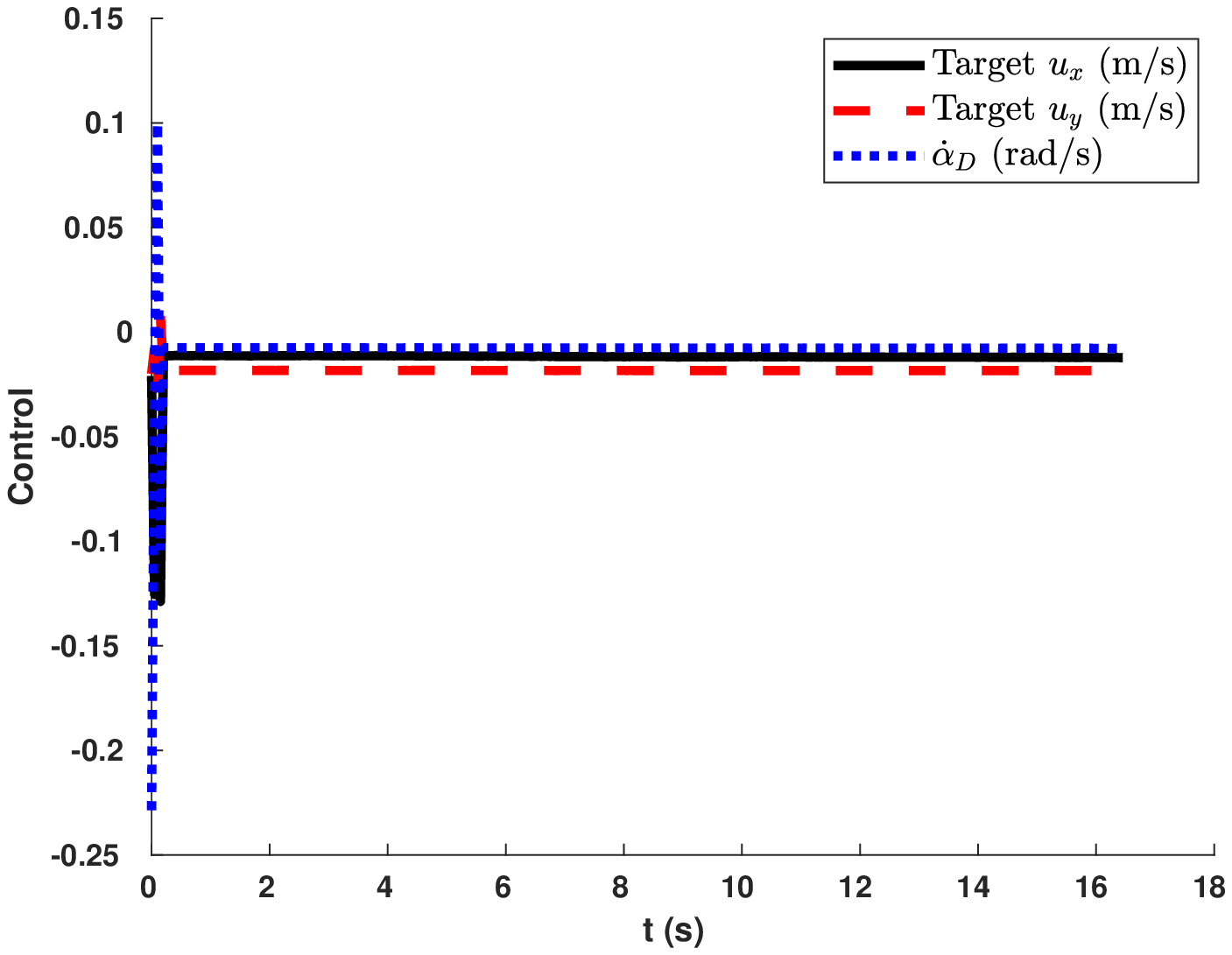}
		\caption{}
		\label{fig:var_escape_control}
	\end{subfigure}
	\begin{subfigure}{6cm}
		\includegraphics[width=\linewidth]{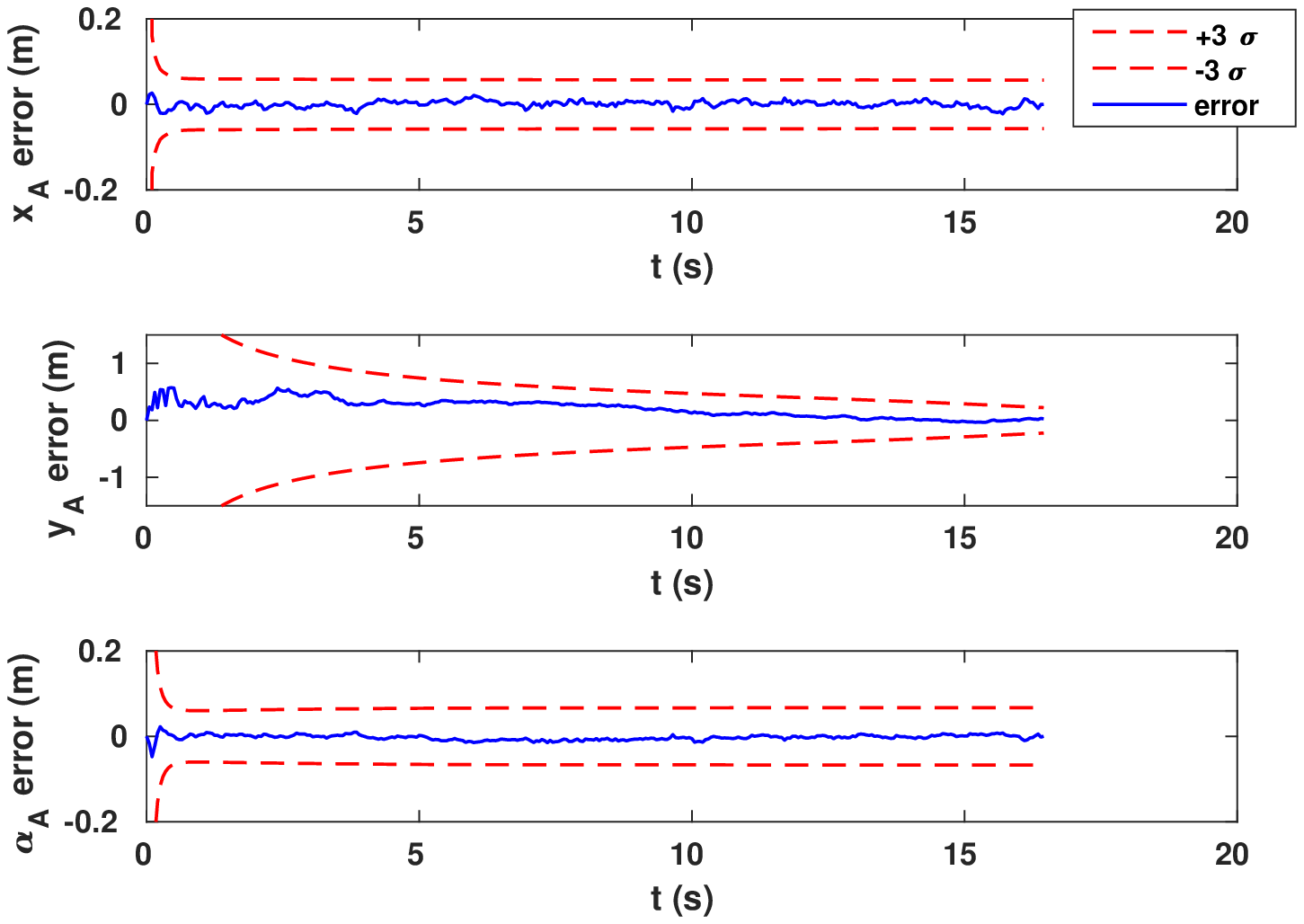}
		\caption{}
		\label{fig:var_escape_error}
	\end{subfigure}
		\caption{Target escape scenario for the variable velocity target where $e$ is not violated. (a) Trajectories of the agents. (b) Distance between the agents. (c) Optimal control inputs determined by the NMPC. (d) Error in the attacker state estimates.}
		\label{fig:var_escape}
\end{figure*}

Target escape scenario for the variable velocity target case, where the safe distance parameter $e$ is not violated is considered. The initial position of the target is selected inside the escape zone, $Z_e'$ and is represented by $T_e(-0.5,1)$ in Fig.~\ref{fig:var_map}. The agent trajectories and the evolution of the distances for this scenario are shown in Fig.~\ref{fig:var_escape_traj} and Fig.~\ref{fig:var_escape_r_R}. It can be seen from the figures that the target is stationary since the attacker was intercepted by the defender before $e$ is violated. It is evident that the target's control inputs would be zero for this case, which is verified in Fig.~\ref{fig:var_escape_control}. The defender's angular velocity is also almost zero due to the near straight line path taken by the agent. Fig.~\ref{fig:var_escape_error} shows the errors in the estimates of the attacker position and heading angle. The errors are low and stay within the $ 3\sigma $ bounds.

\subsubsection{Target escape case where safe distance is violated}
\begin{figure*}
	\centering 
	\begin{subfigure}{6cm}
		\includegraphics[width=\linewidth]{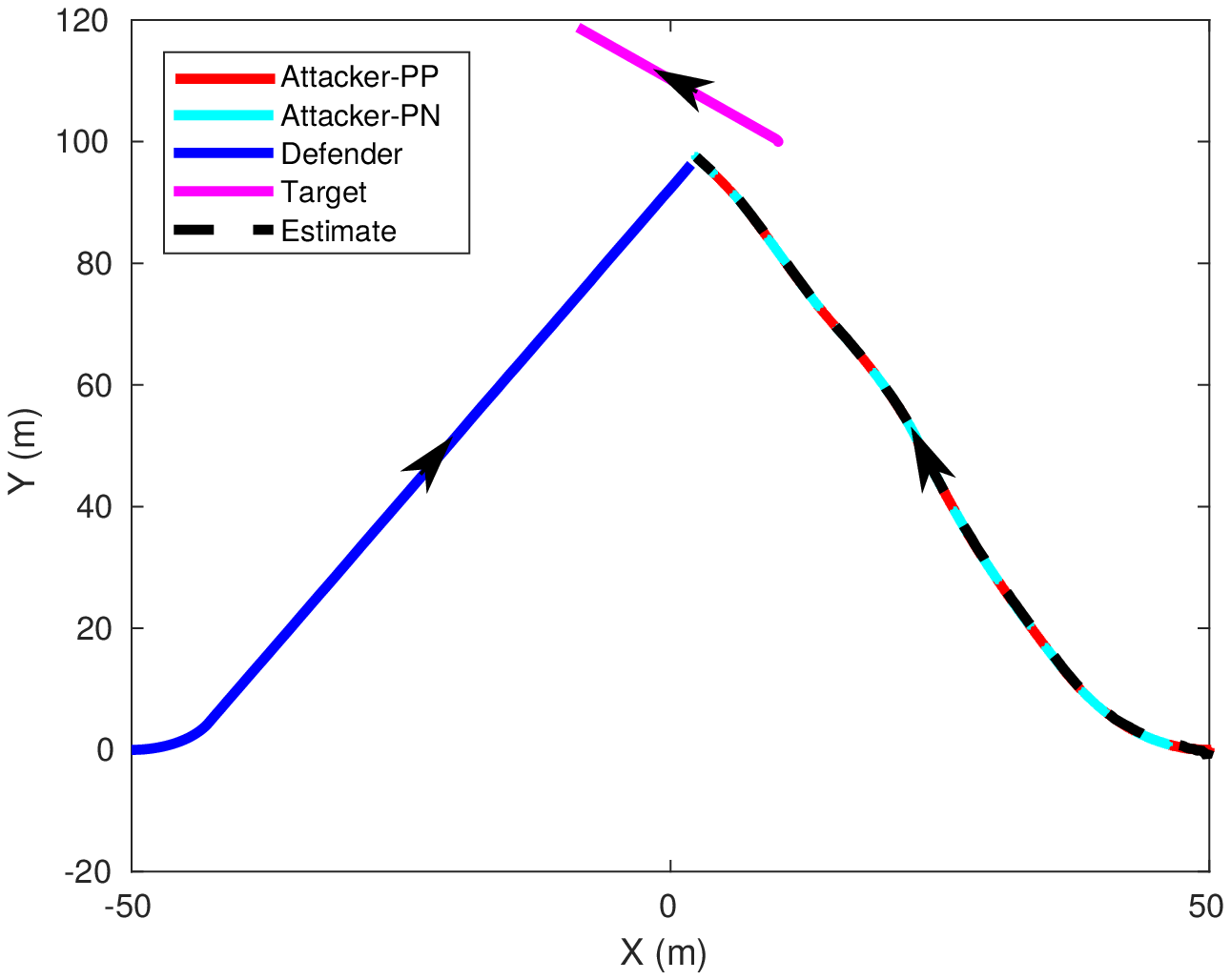}
		\caption{}
		\label{fig:var_escape_c3_traj}
	\end{subfigure}
	\begin{subfigure}{6cm}
		\includegraphics[width=\linewidth]{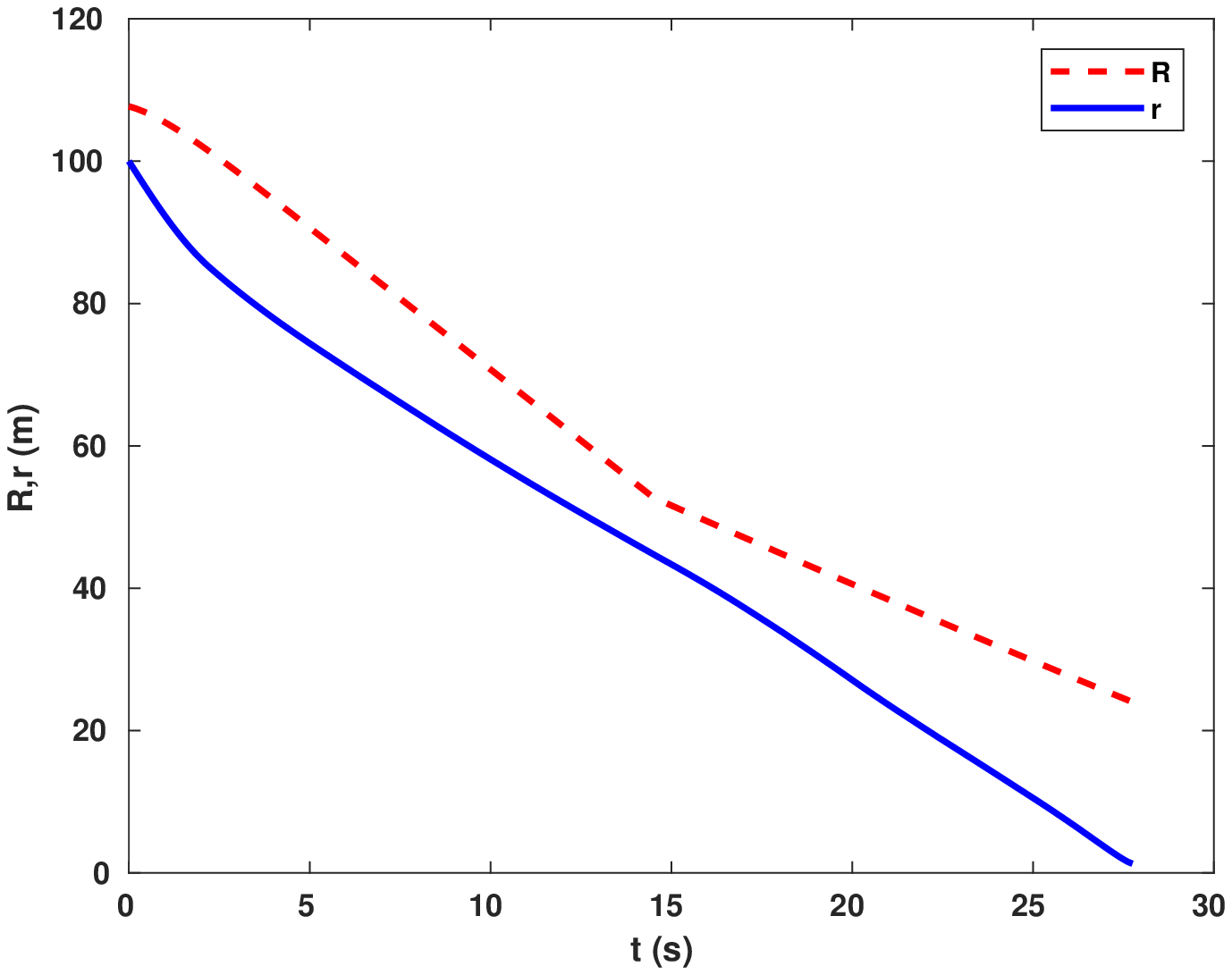}
		\caption{}
		\label{fig:var_escape_c3_r_R}
	\end{subfigure}
    
	\begin{subfigure}{6cm}
		\includegraphics[width=\linewidth]{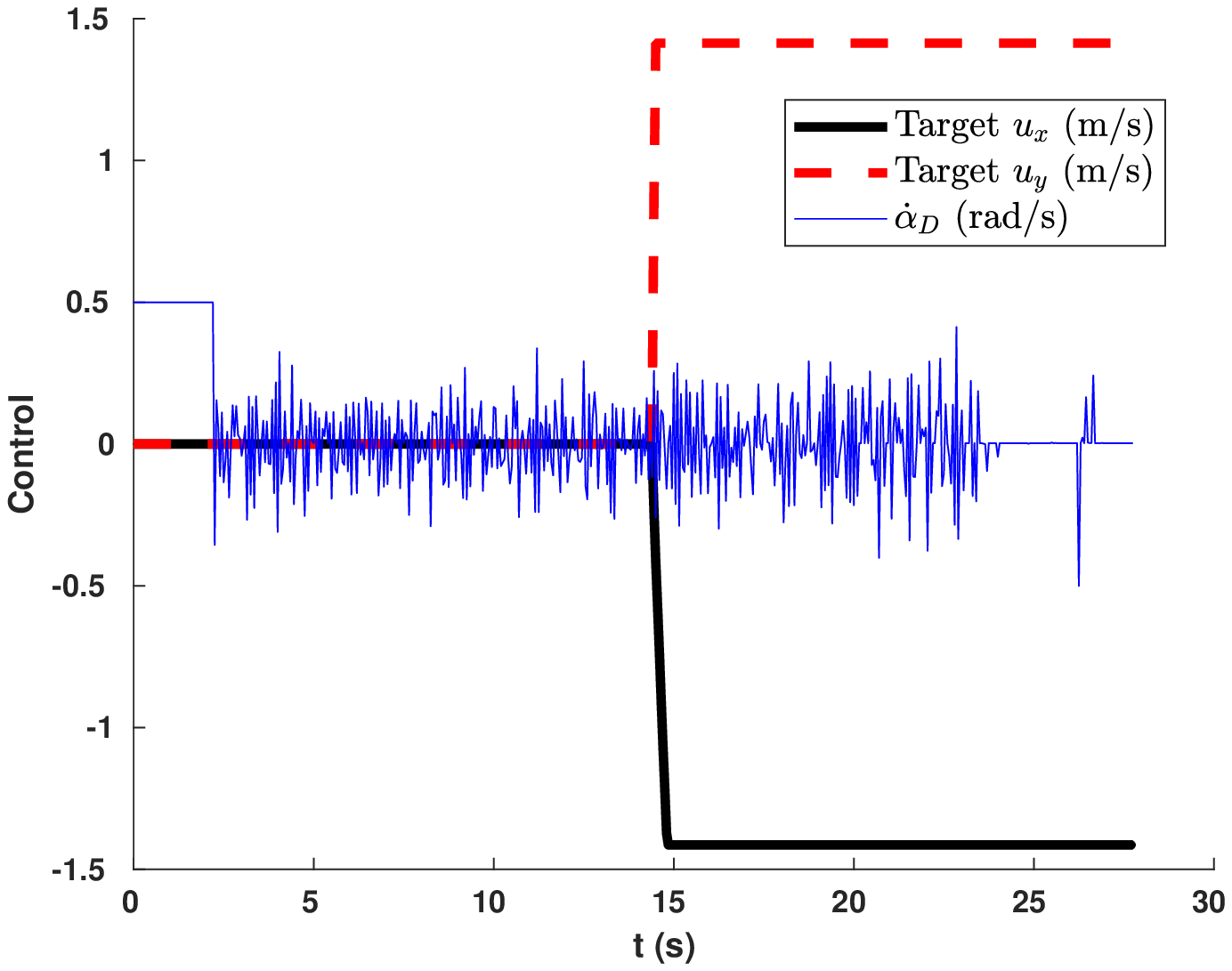}
		\caption{}
		\label{fig:var_escape_c3_control}
	\end{subfigure}
	\begin{subfigure}{6cm}
		\includegraphics[width=\linewidth]{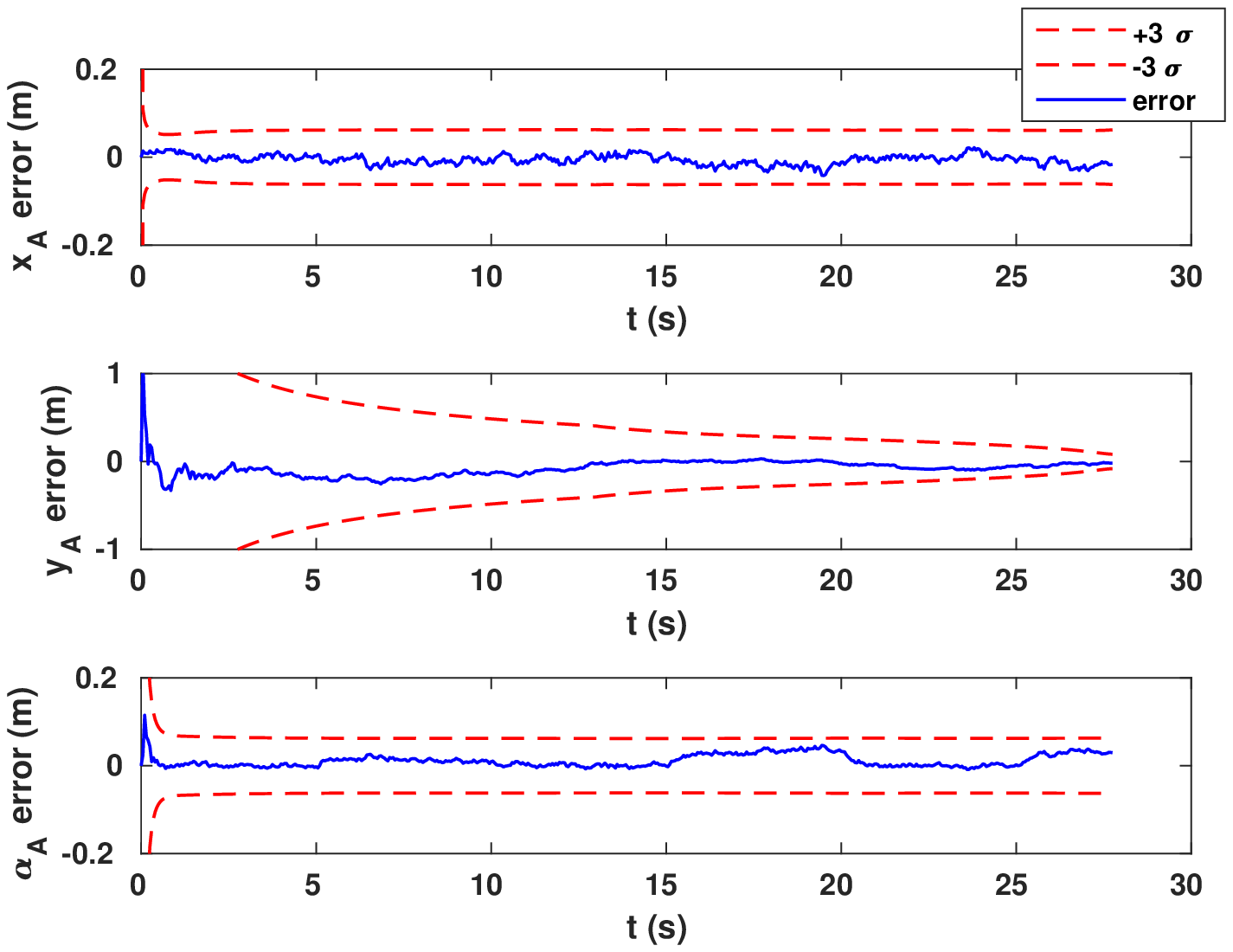}
		\caption{}
		\label{fig:var_escape_c3_error}
	\end{subfigure}
		\caption{Target escape scenario for the variable velocity target where $e$ is violated. (a) Trajectories of the agents. (b) Distance between the agents. (c) Optimal control inputs determined by the NMPC. (d) Error in the attacker state estimates.}
		\label{fig:var_escape_c3}
\end{figure*}

Target escape scenario for the variable velocity target case, where the safe distance parameter $e$ is violated is considered here. The initial position of the target is selected inside the escape zone, $Z_e'$ and is represented by $T_{ee}(0.1,1)$ in Fig.~\ref{fig:var_map}. The agent trajectories and the evolution of the distances for this scenario are shown in Fig.~\ref{fig:var_escape_c3_traj} and Fig.~\ref{fig:var_escape_c3_r_R}. It can be seen from the figures that the target started moving when $e$ was about to be violated, and the target's control inputs became nonzero as shown in Fig.~\ref{fig:var_escape_c3_control}. It can be seen from Fig.~\ref{fig:var_escape_c3_r_R} that the slope of $ R $ was decreasing at a constant rate, and the rate decreased when the target started moving. Fig.~\ref{fig:var_escape_c3_error} shows the errors in the estimates of the attacker position and heading angle, which is very low and stays within the 3$\sigma$ bounds.

\subsubsection{Target capture case}
\begin{figure*}
	\centering 
	\begin{subfigure}{5.5cm}
		\includegraphics[width=5cm]{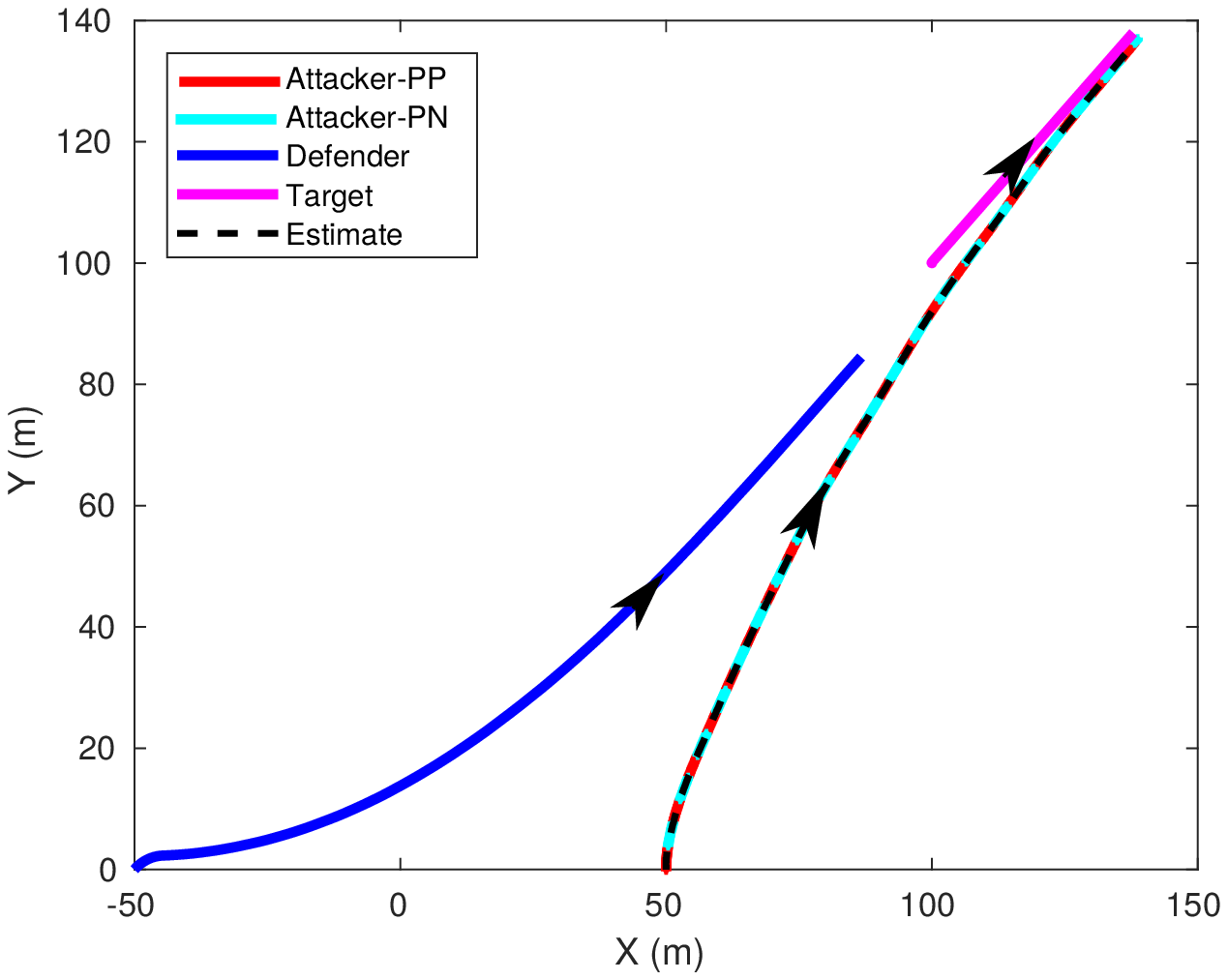}
		\caption{}
		\label{fig:var_capture_traj}
	\end{subfigure}
	\begin{subfigure}{5.5cm}
		\includegraphics[width=5cm]{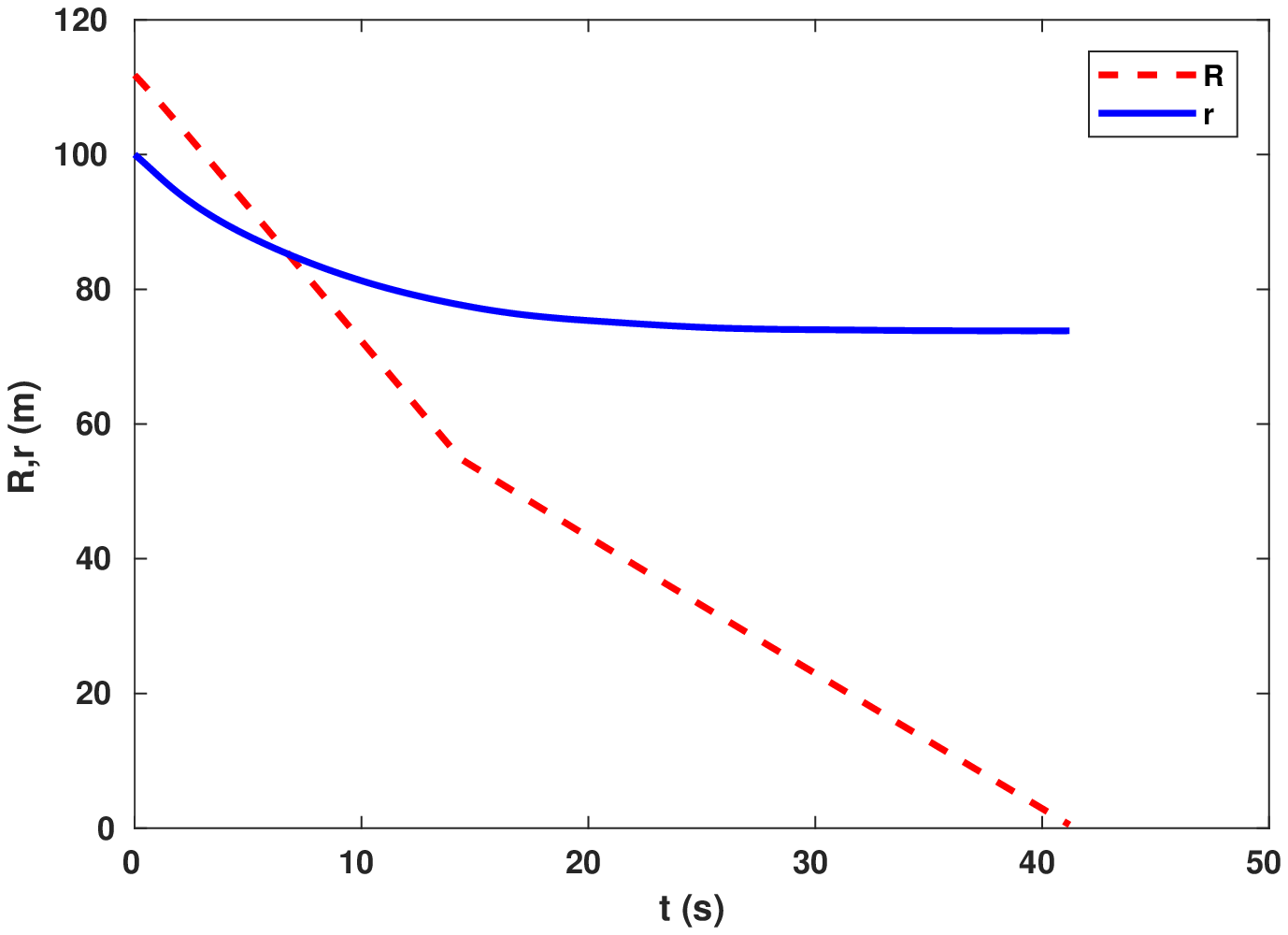}
		\caption{}
		\label{fig:var_capture_r_R}
	\end{subfigure}
	\begin{subfigure}{5.5cm}
		\includegraphics[width=5cm]{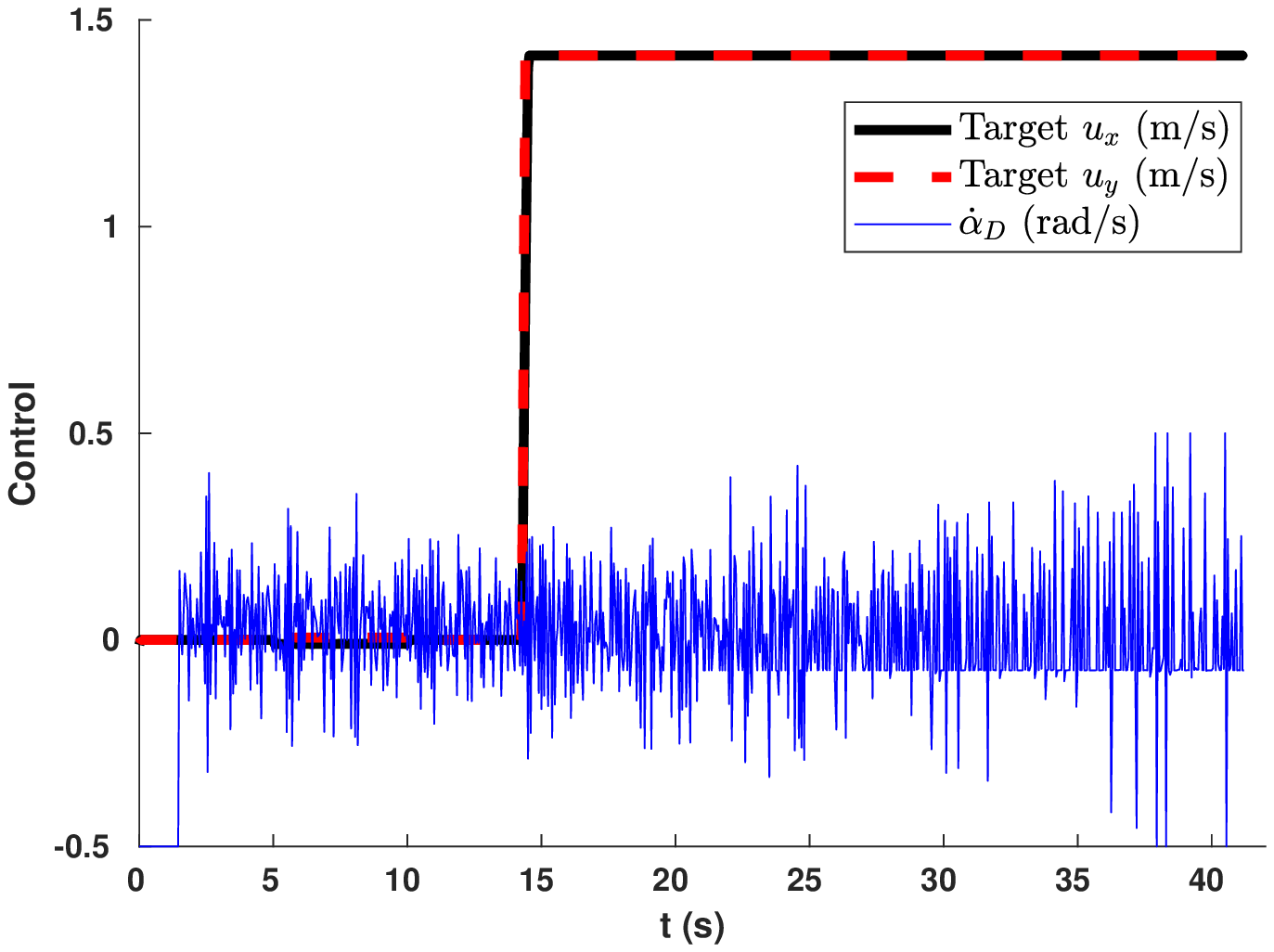}
		\caption{}
		\label{fig:var_capture_control}
	\end{subfigure}
		\caption{Target capture scenario for the variable velocity target. (a) Trajectories of the agents. (b) Distance between the agents. (c) Optimal control inputs determined by the NMPC. 
		}
		\label{fig:var_capture}
\end{figure*}

The initial position of the target is selected inside the capture zone, $Z_c'$ and is represented by $T_c(1,1)$ in Fig.~\ref{fig:var_map}. The agent trajectories and the evolution of distances between the agents are shown in Fig.~\ref{fig:var_capture_traj} and Fig.~\ref{fig:var_capture_r_R} respectively. The optimal control inputs determined by the NMPC for the target and the defender are shown in Fig.~\ref{fig:var_capture_control}. It can be seen that the target's control inputs become nonzero only after the safe distance $e$ was about to be violated. The target then moves with the maximum allowed velocity and tries to maintain $R>e$. The target was captured as the defender failed in intercepting the attacker. Similar to the other simulations, the estimated errors were below the 3$\sigma$ bounds. 

\subsection{Unequal attacker--defender speed ratio case}
\begin{figure}[h]
	\centering
	\includegraphics[width=6cm]{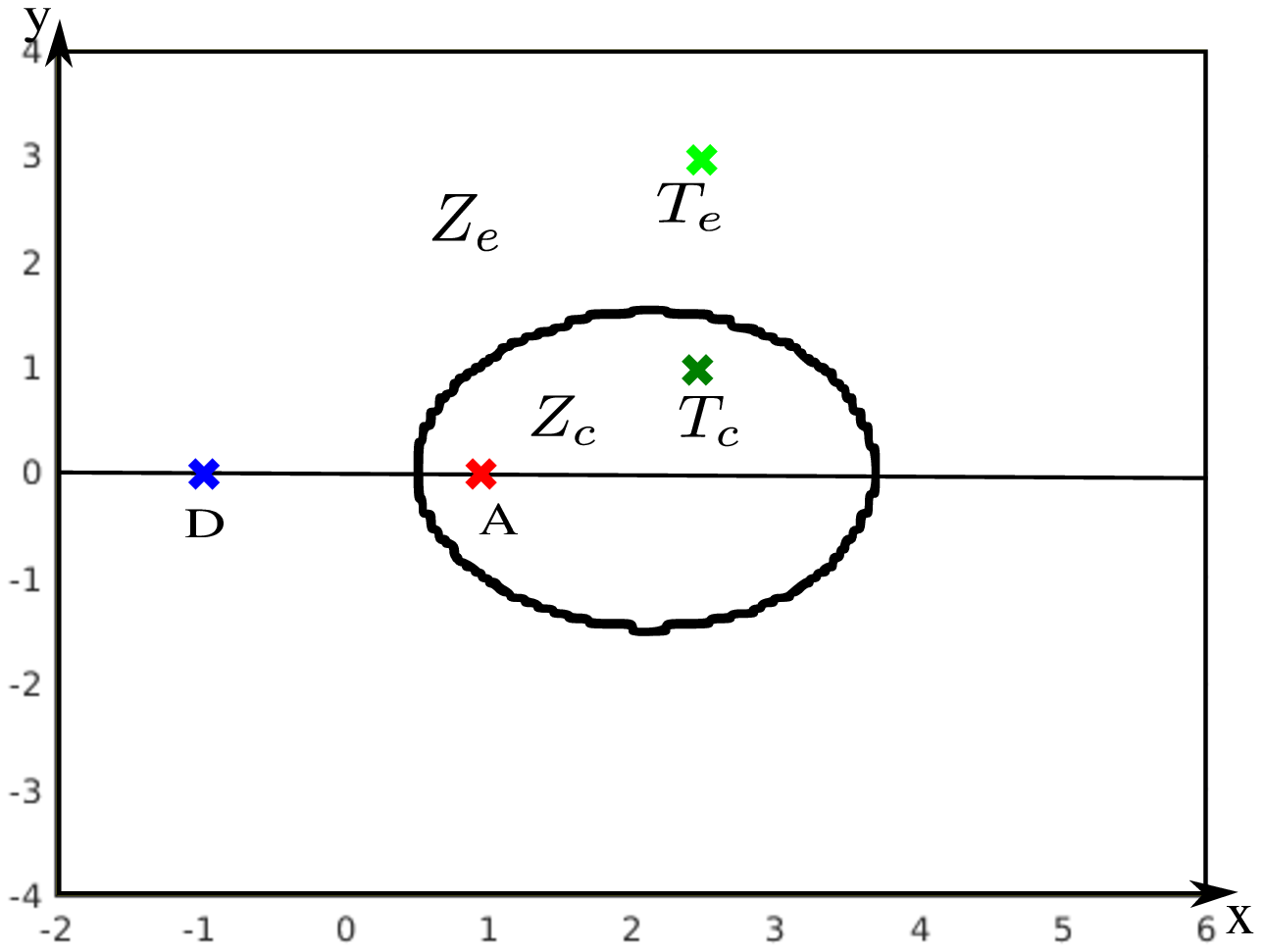}
	\caption{Initial agent configurations for the variable velocity target case, $\gamma_{AD}=1.5$. }
	\label{fig:variable_gamma2_map}
\end{figure}

Simulations were carried out with different configurations (constant speed and variable velocity target, $A-D$ speed ratio $\gamma_{AD}<1$ and $\gamma_{AD}>1$) and only one example case is presented here to avoid redundancy and respect the space constraints. Consider the initial configuration for a variable target velocity case given in Fig.~\ref{fig:variable_gamma2_map}. The initial parameters of the agents are given in Table~\ref{table:initial_unequal}. The $A-D$ speed ratio is selected as $\gamma_{AD}=1.5$ and the $A-T$ speed ratio, $\gamma_{AT}=0.5$. The safe distance parameter $e$ is selected as half of the initial distance between the target and the attacker.

\begin{table}
	\centering
	\begin{tabular}{ |c|c|c| } 
		\hline
		\multicolumn{3}{|c|}{Unequal $A-D$ speed ratio}\\
		\multicolumn{3}{|c|}{(variable velocity target)}\\
		\hline 
		parameter&escape&capture\\
		\hline
		$ (x_{A},y_{A}) $ (m) & (10,0) & (10,0) \\ 
		$ (x_{T},y_{T}) $ (m) & (25,30) & (25,10) \\
		$ (x_{D},y_{D}) $ (m) & (-10,0) & (-10,0) \\
		$ \alpha_{A} $ (rad)  &  0.78   &  0.78 \\ $u_{x} $ (m/s)  &  0 & 0 \\
		$u_{y} $ (m/s)  &  0 & 0 \\
		$\alpha_{D} $ (rad)  &  0   &  0 \\
		$ v_{A} $ (m/s)  &  4   &   4 \\
		$ v_{D} $ (m/s)  &  6   &   6  \\
		\hline
	\end{tabular}
	\caption{Initial parameters for the agents for unequal $A-D$ speed ratio case.}
	\label{table:initial_unequal}
\end{table}

\subsubsection{Target escape case}
\begin{figure*}[h]
	\centering 
	\begin{subfigure}{5.5cm}
		\includegraphics[width=\linewidth]{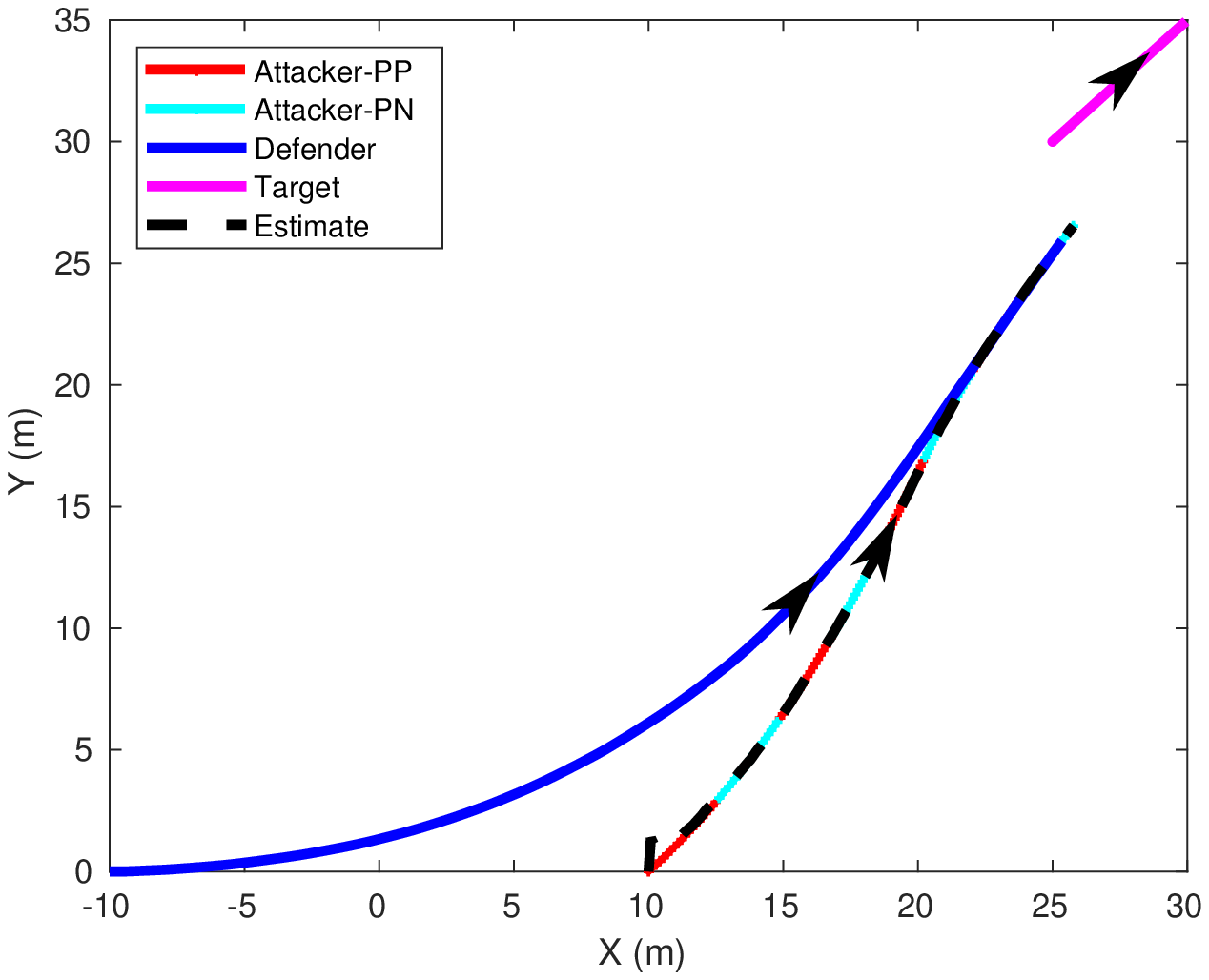}
		\caption{}
		\label{fig:vv_gamma2_escape_traj}
	\end{subfigure}
	\begin{subfigure}{5.5cm}
		\includegraphics[width=\linewidth]{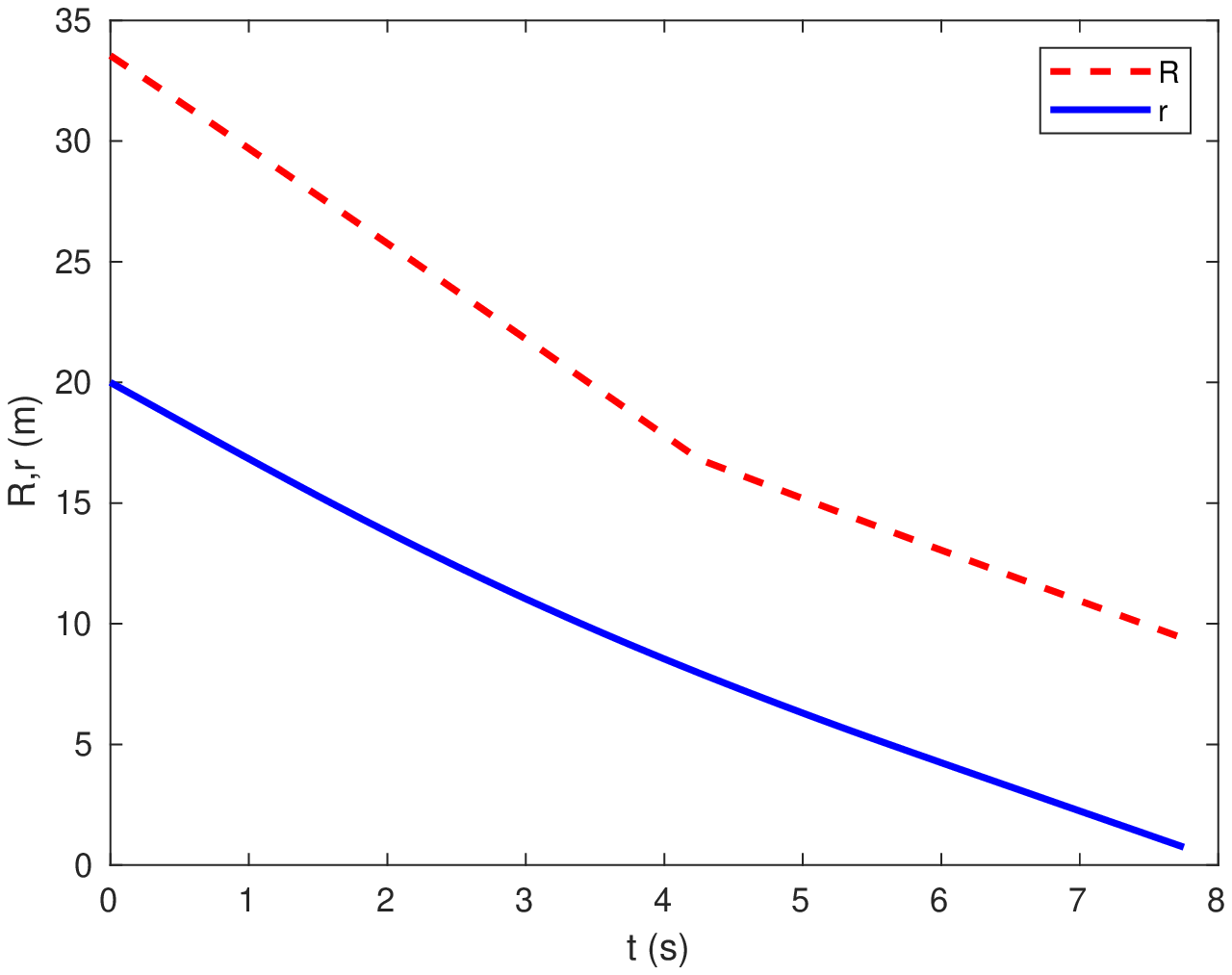}
		\caption{}
		\label{fig:vv_gamma2_escape_rR}
	\end{subfigure}
	\begin{subfigure}{5.5cm}
		\includegraphics[width=\linewidth]{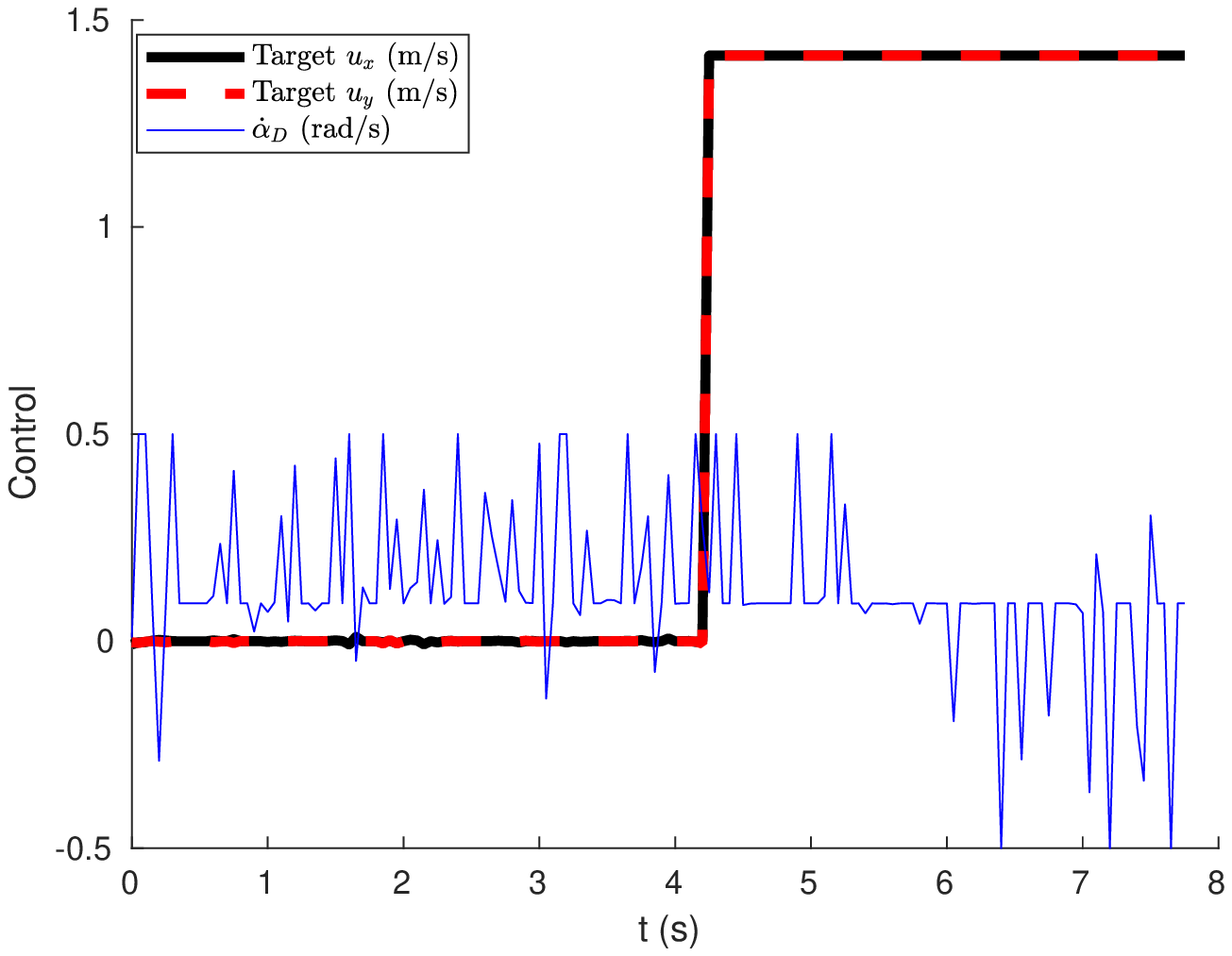}
		\caption{}
		\label{fig:vv_gamma2_escape_control}
	\end{subfigure}
	\caption{Target escape scenario for the variable velocity target, $\gamma_{AD}=1.5$ (a) Trajectories of the agents. (b) Distance between the agents. (c) Optimal control inputs determined by the NMPC. 
	}
	\label{fig:vv_gamma2_escape}
\end{figure*}

The initial position of the target is selected inside the escape zone, $Z_e$ and is represented by $T_{e}(2.5,3)$ in Fig.~\ref{fig:variable_gamma2_map}. The agent trajectories and the evolution of the distances for this scenario are shown in Fig.~\ref{fig:vv_gamma2_escape_traj} and Fig.~\ref{fig:vv_gamma2_escape_rR}. The safe distance $e$ was violated, and hence the target's control inputs became nonzero as shown in Fig.~\ref{fig:vv_gamma2_escape_control}. The distance between the attacker and the defender goes to zero in Fig.~\ref{fig:vv_gamma2_escape_rR} confirming the $A-D$ interception.

\subsubsection{Target capture case}
\begin{figure*}[h]
	\centering 
	\begin{subfigure}{5.5cm}
		\includegraphics[width=5cm]{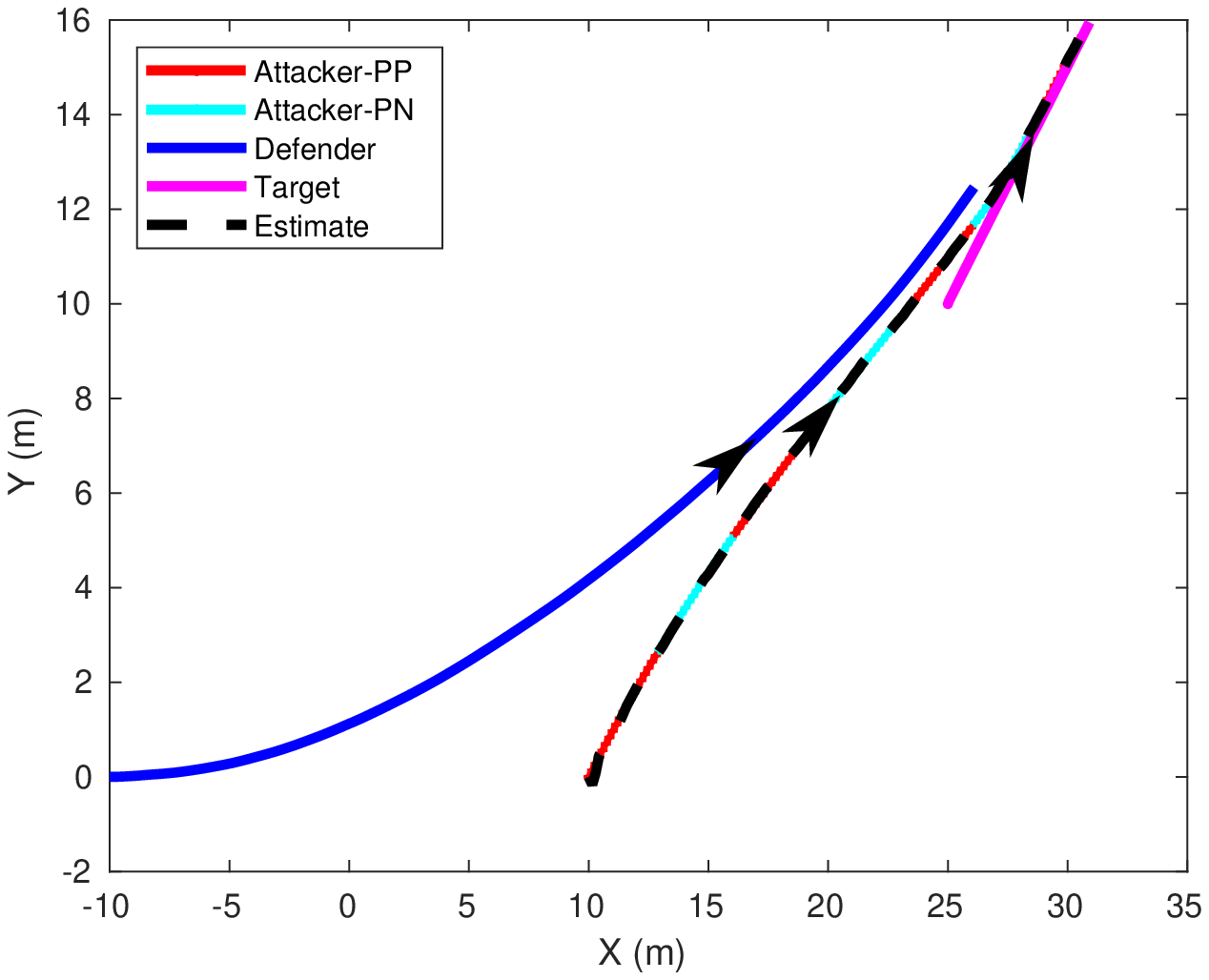}
		\caption{}
		\label{fig:vv_gamma2_capture_traj}
	\end{subfigure}
	\begin{subfigure}{5.5cm}
		\includegraphics[width=5cm]{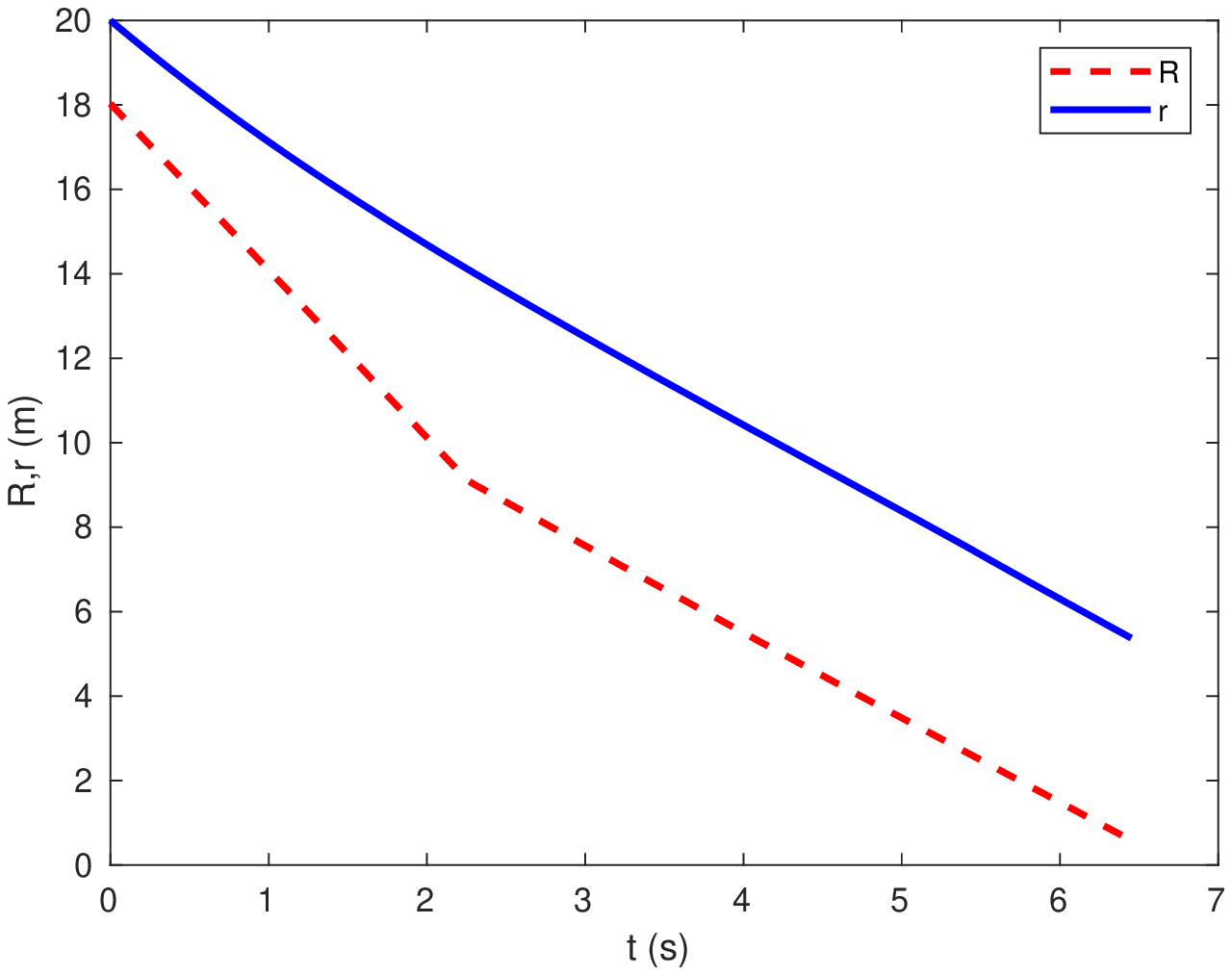}
		\caption{}
		\label{fig:vv_gamma2_capture_rR}
	\end{subfigure}
	\begin{subfigure}{5.5cm}
		\includegraphics[width=5cm]{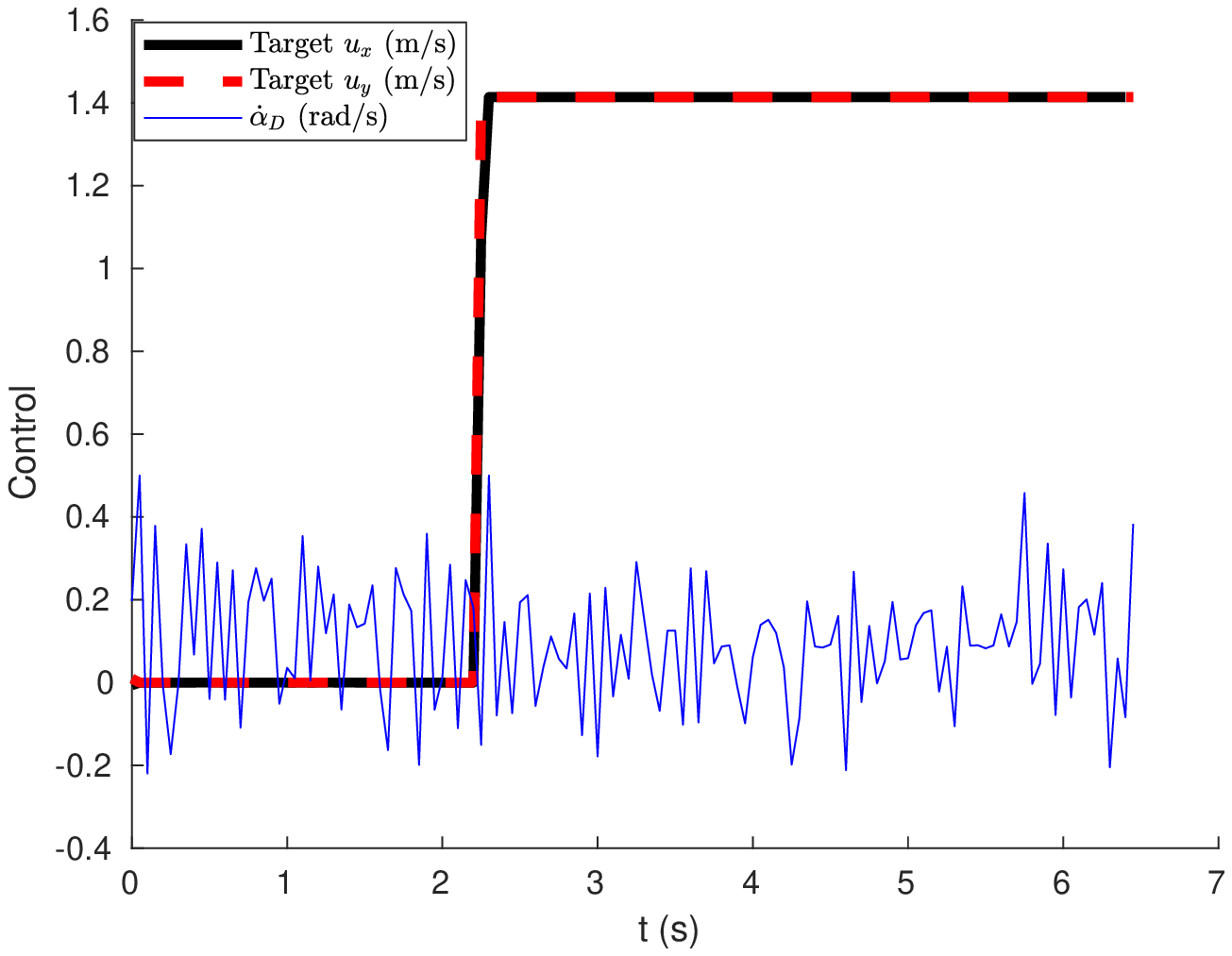}
		\caption{}
		\label{fig:vv_gamma2_capture_control}
	\end{subfigure}
	\caption{Target capture scenario for the variable velocity target, $\gamma_{AD}=1.5$. (a) Trajectories of the agents. (b) Distance between the agents. (c) Optimal control inputs determined by the NMPC. 
	}
	\label{fig:vv_gamma2_capture}
\end{figure*}

The initial position of the target is selected inside the capture zone, $Z_c$ and is represented by $T_c(2.5,1)$ in Fig.~\ref{fig:variable_gamma2_map}. The agent trajectories and the evolution of distances between the agents are shown in Fig.~\ref{fig:vv_gamma2_capture_traj} and Fig.~\ref{fig:vv_gamma2_capture_rR} respectively. The optimal control inputs determined by the NMPC for the target and the defender are shown in Fig.~\ref{fig:vv_gamma2_capture_control}. Even though the control inputs became nonzero after the safe distance $e$ was violated, the target failed to escape since the defender was unable to intercept the attacker.

\begin{figure}
	\centering
	\includegraphics[width=6cm]{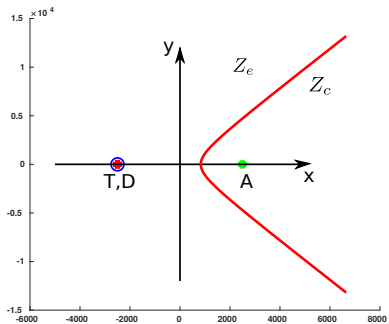}
	\caption{Initial agent configurations for the comparison of CLOS, A-CLOS, and NMPC. }
	\label{fig:clos_aclos_map}
\end{figure}

\begin{figure*}[h]
	\centering 
	\begin{subfigure}{6cm}
		\includegraphics[width=6cm]{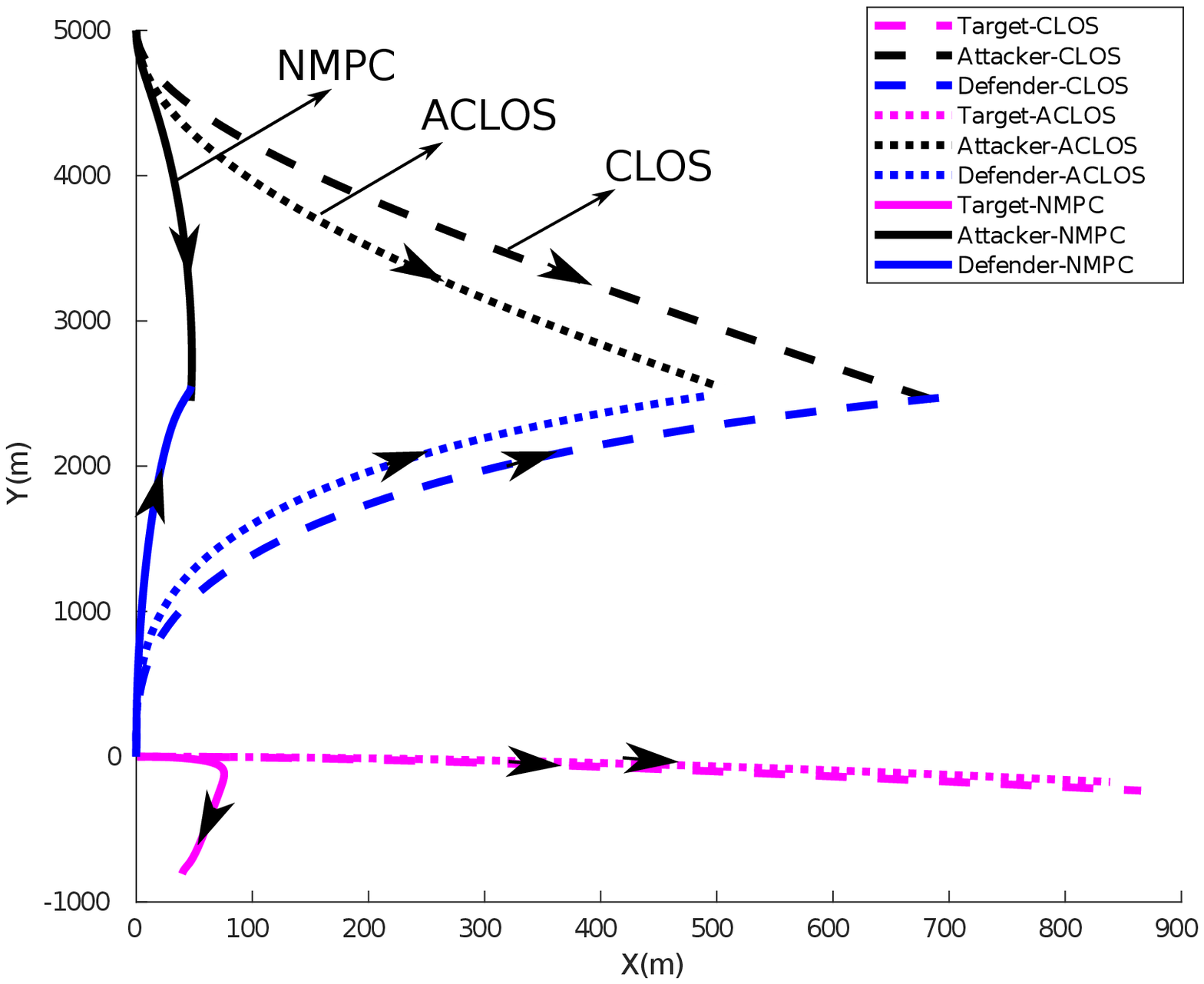}
		\caption{}
		\label{fig:clos_aclos_mpc_traj}
	\end{subfigure}
	\begin{subfigure}{6cm}
		\includegraphics[width=6cm]{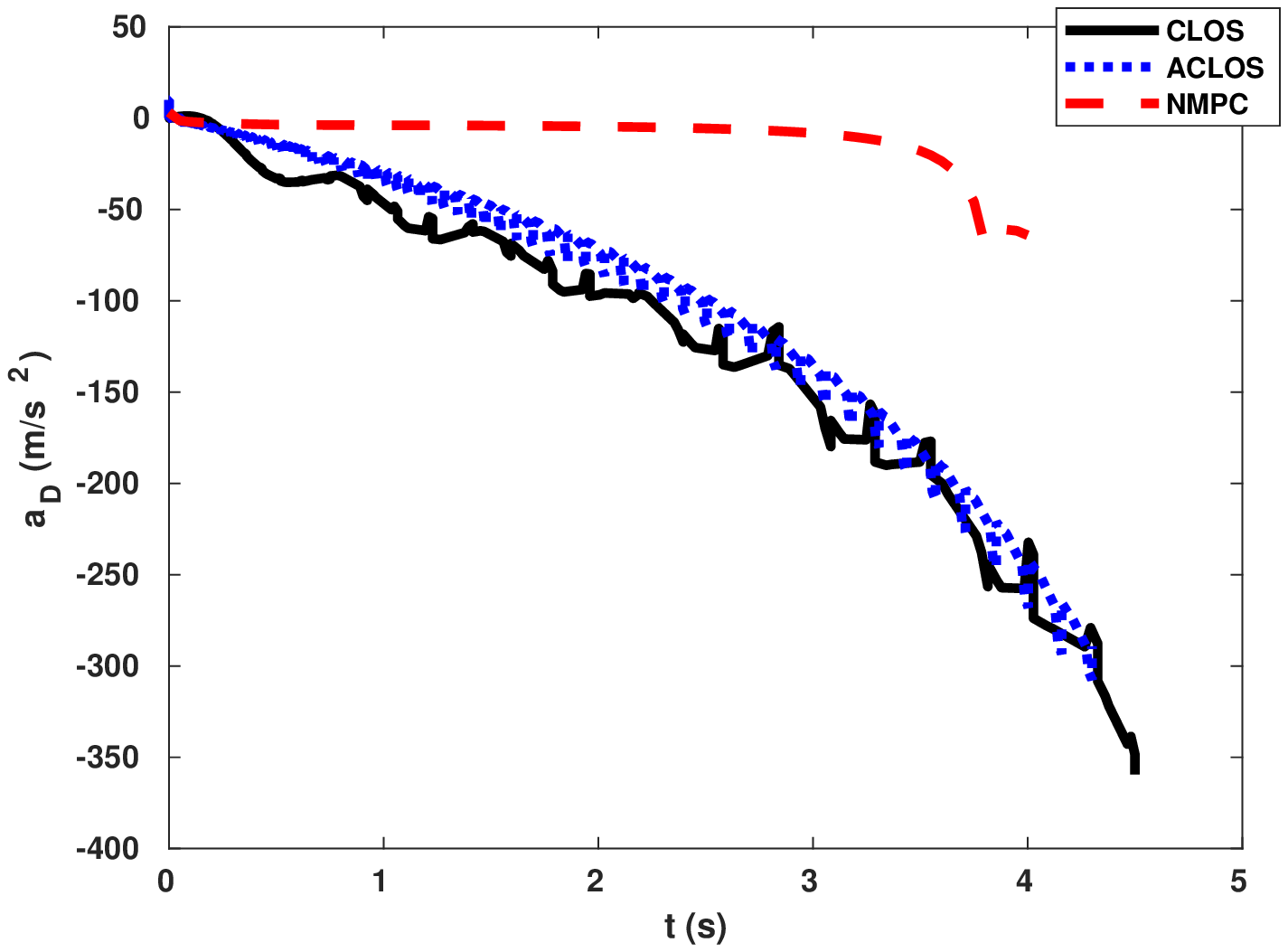}
		\caption{}
		\label{fig:clos_aclos_mpc_control}
	\end{subfigure}
		\caption{Performance comparison of the NMPC, CLOS and the A-CLOS. (a) Agent trajectories with CLOS, A-CLOS, and NMPC. (b) Control effort comparison for the CLOS, A-CLOS, and the NMPC.}
		\label{fig:clos_aclos_mpc}
\end{figure*}
\subsection{Comparison with CLOS and A-CLOS guidance law}
The proposed NMPC scheme for the three-agent pursuit-evasion problem was compared with existing solutions which use a command to line-of-sight (CLOS) guidance~\cite{ratnoo2011line} and a modified CLOS guidance, called A-CLOS law~\cite{yamasaki2013modified}, for the calculation of control input for the defender. The three agent system was simulated with the same initial conditions as given in Table~\ref{table:initial_comparison} and was compared with the solutions given by the CLOS and the A-CLOS guidance laws.


\begin{table}
	\centering
	\begin{tabular}{ |l|c|c|c|}		\hline 
		
		Parameter&NMPC&A-CLOS&CLOS\\
		\hline
		interception time (s) & 4.0 & 4.3&4.5 \\
		avg. control effort (m/s$^2$) & 350.97 & 3.38e+03 & 1.11e+04\\
		avg. computation time / iteration & 0.0374 & 0.0018 & 0.0038 \\
		\hline
	\end{tabular}
	\caption{Performance comparison between NMPC, A-CLOS, and CLOS.}
	\label{table:initial_comparison}
\end{table}

Velocities of the agents were selected as $ v_{T}=200 $\,m/s, $ v_{A}=600 $\,m/s and $ v_{D}=600 $\,m/s respectively to match the parameters given in~\cite{ratnoo2011line}. The initial positions of the agents are shown in Fig.~\ref{fig:clos_aclos_map}. The target and the defender starts from the same initial position, and the target moves with constant speed. The initial position of the target lies in the escape zone $Z_e$. The agent trajectories for the CLOS, A-CLOSG and the NMPC for the selected scenario is given in Fig.~\ref{fig:clos_aclos_mpc_traj}. The computed control input for the defender, which is the lateral acceleration, is given in Fig.~\ref{fig:clos_aclos_mpc_control} for all the approaches. The simulation results show that the NMPC outperformed the CLOS and A-CLOS with lower interception time and lower average control effort as seen from Table~\ref{table:initial_comparison}. Even though the NMPC has a higher computation time compared to the CLOS and A-CLOS, it should be considered as a trade-off between computation time and efficiency. Nevertheless, the proposed NMPC scheme is real-time implementable according to our experience~\cite{manoharan2019nmpc}. 
\section{Conclusions}\label{sec:conclusions}
A nonlinear model predictive control strategy was proposed for the active defense of the target in a TAD game. The formulation involves computing control commands for a cooperative target-defender pair against an individually acting attacker. The state information of the attacker was assumed to be unknown and was estimated using an EKF. An analysis using the Apollonius circles was conducted to determine the escape regions for the target, and the same was verified using simulations conducted for various scenarios. The nonlinear online feedback scheme designed using the NMPC was found to be effective in achieving the objectives while respecting the imposed constraints. The performance of the NMPC strategy was compared against CLOS and A-CLOS, and the results showed that the NMPC outperformed both the CLOS and A-CLOS based strategies. 

Possible extensions in this area of study are to broaden the framework to three dimensions taking the terrain map into account and including obstacles for the three-agent TAD game in urban environments. Also, we would like to drive our approach towards the use of fast-MPC based solvers for speeding up the computations in the proposed strategy.


 	\bibliographystyle{IEEEtran}
	\bibliography{ref}
\end{document}